\documentclass[aps,showkeys,nofootinbib,superscriptaddress,twocolumn]{revtex4-2}

\usepackage{enumerate}
\usepackage{amsmath,slashed,tensor,amssymb,epsfig,amsthm,bm,graphicx,graphics,slashed}
\usepackage{xcolor}
\usepackage[utf8]{inputenc}
\usepackage[top=1.5cm, bottom=1.5cm, left=2cm, right=2cm]{geometry}
\usepackage{lipsum}
\usepackage{mathtools}
\usepackage{graphicx}
\usepackage{subcaption}
\usepackage{braket}
\usepackage{dsfont}
\usepackage{orcidlink}
\usepackage[justification=raggedright,singlelinecheck=false]{caption}

\makeindex

\begin{document}

\title{Uncertainty equality for SU($N$) observables enabling the experimentally friendly detection of $k$-inseparability via purity measurements}

\author{Gianluigi Tartaglione
\orcidlink{0009-0003-1594-4733}}

\affiliation{Dipartimento di Ingegneria Industriale, Università degli Studi di Salerno, Via Giovanni Paolo II 132, 84084 Fisciano (SA), Italy}

\affiliation{Institute of Nanotechnology of the National Research Council of Italy, CNR-NANOTEC, Lecce Central Unit, c/o Campus Ecotekne, Via Monteroni, 73100 Lecce, Italy}

\author{Gennaro Zanfardino
\orcidlink{0009-0002-4202-6498}}

\affiliation{Dipartimento di Ingegneria Industriale, Università degli Studi di Salerno, Via Giovanni Paolo II 132, 84084 Fisciano (SA), Italy}

\affiliation{Institute of Nanotechnology of the National Research Council of Italy, CNR-NANOTEC, Lecce Central Unit, c/o Campus Ecotekne, Via Monteroni, 73100 Lecce, Italy}

\affiliation{Dipartimento di Medicina Sperimentale, Università del Salento, c/o Campus Ecotekne, Via Monteroni, 73100 Lecce, Italy}

\affiliation{INFN, Sezione di Napoli, Gruppo Collegato di Salerno, Italy}

\author{Fabrizio Illuminati
\orcidlink{0000-0002-2273-0421}}
\affiliation{Institute of Nanotechnology of the National Research Council of Italy, CNR-NANOTEC, Lecce Central Unit, c/o Campus Ecotekne, Via Monteroni, 73100 Lecce, Italy}

\affiliation{Dipartimento di Ingegneria Industriale, Università degli Studi di Salerno, Via Giovanni Paolo II 132, 84084 Fisciano (SA), Italy}

\affiliation{INFN, Sezione di Napoli, Gruppo Collegato di Salerno, Italy}

\date{\today}

\begin{abstract}

We derive an exact uncertainty relation for arbitrary quantum states of finite-dimensional Hilbert spaces. For any given $k$-partition of a $d$-dimensional multipartite system, we introduce the \textit{total uncertainty} as the sum of the uncertainties associated with all possible tensor products of local $\mathrm{SU}(N)$ observables, where each observable acts on the corresponding subsystem.
We show that the total uncertainty exactly equals the algebraic sum of the global state purity and the purities of all possible state reductions, embodying an uncertainty conservation law: at fixed purities, any redistribution of local uncertainties leaves the total uncertainty invariant. 
For systems containing at least one single-qubit subsystem, this equality implies saturation of the Robertson-Schr\"odinger uncertainty inequality, with the missing term needed for saturation equal to the bipartite qubit-environment entanglement for a pure global state, or to the qubit two-Rényi entropy for a mixed global state. Leveraging on these results, we show how for any finite-dimensional multipartite system the Hilbert-Schmidt squared norm of the system correlation matrix $t$ can be expressed exclusively in terms of the global and reduced state purities or, equivalently, in terms of the global and reduced state two-Rényi entropies. We then derive a correlation matrix-based necessary condition for $k$-separability of arbitrary finite-dimensional quantum states and show, in the case of $n$ qubits, how it is related to a necessary criterion for Bell nonlocality in scenarios with two dichotomic measurements per party. 
Since the number of global and reduced state purities always scales as $2^k$ irrespective of the local Hilbert space dimension, for sufficiently large systems the purity-based formulation of the $k$-separability criterion always yields an exponential advantage over the direct evaluation of the $t$-matrix norm, allowing for a more efficient practical verification of multipartite entanglement and nonlocality via simple experimental schemes based on purity measurements or, equivalently, on the measurement of the state two-Rényi entropies. Our results shed some further light on the intimate and intricate relation between correlations, entropies, uncertainties, and the entanglement certification and detection problem.

\end{abstract}

\maketitle

\section{Introduction}

One of the main structural traits of quantum mechanics is the existence of fundamental limits to the precision with which certain pairs of incompatible physical observables, can be simultaneously measured. These limitations are expressed through the well known \textit{uncertainty relations}.

Uncertainty relations are not merely limitations of experimental techniques; rather, they are inherent to the quantum description of nature. Originally introduced by Heisenberg~\cite{Heisenberg1927,Wheeler2014}, they arise from the non-commutativity of the  operators associated with the corresponding observables. For any pair of observables $\hat{A}$ and $\hat{B}$, and any quantum state $\rho$, the Robertson uncertainty relation generalizes Heisenberg's result originally stated for position $\hat{x}$ and momentum $\hat{p}$~\cite{Robertson1929}:
\begin{equation}
\Delta^2 A \, \Delta^2 B \geq  \left|\frac{1}{2} \langle [\hat{A}, \hat{B}] \rangle \right|^2,
\label{uncertainty_HR}
\end{equation}
where $[\hat{A}, \hat{B}] = \hat{A}\hat{B} - \hat{B}\hat{A}$ is the commutator, for any observable $\hat{\mathcal{O}}$, $\Delta ^2\mathcal{O}=\langle \hat{\mathcal{O}^2}\rangle-\langle \hat{\mathcal{O}} \rangle^2$, and $\langle \cdot \rangle$ denotes the expectation value with respect to state $\rho$.

Inequality~\eqref{uncertainty_HR} was then extended by Schrödinger to the general form that goes under the name of Robertson-Schrödinger uncertainty relation~\cite{Schrodinger1930}: 

\begin{equation}
    \Delta^2 A  \Delta^2 B =  \left|\frac{1}{2} \langle [\hat{A}, \hat{B}] \rangle \right|^2 + \left|  \frac{1}{2} \langle \{ \hat{A}, \hat{B} \} \rangle - \langle \hat{A} \rangle\langle \hat{B} \rangle\right|^2 + \mathcal{R} \, ,
    \label{equality_scrhrodinger_unciertainty}
 \end{equation}
where the anti-commutator $\{ \hat{A},\hat{B} \}=\hat{A}\hat{B} + \hat{B}\hat{A}$. The unknown missing term $\mathcal{R}$ needed to satisfy the equality is usually referred to in the literature as the ``rest''~\cite{CarruthersNieto}.

The quest to further extend the uncertainty relations Eqs.~\eqref{equality_scrhrodinger_unciertainty} to more general scenarios that include the sum of uncertainties associated with incompatible observables has met remarkable success with the discovery of the Maccone-Pati uncertainty relations~\cite{Pati2007,Maccone2014}.
The first Maccone-Pati inequality reads:
\begin{equation}
\Delta^2 A + \Delta^2 B \geq \pm i \langle [\hat A, \hat B] \rangle + \left| \bra{\psi} \hat A \pm i \hat B \ket{\psi^\perp} \right|^2\, ,
\label{eq_maccone_pati_1}
\end{equation}
where $\ket{\psi^{\perp}}$ is an arbitrary state orthogonal to $\ket{\psi}$, and the sign $\pm$ is chosen such that the (real) quantity $\pm i \langle [A, B] \rangle$ is positive.

\vspace{1em}

The second inequality, which provides a nontrivial bound even when $\ket{\psi}$ is an eigenstate of either $\hat A$ or $\hat B$, is given by
\begin{equation}
\Delta^2 A + \Delta^2 B \geq \frac{1}{2} \left| \bra{\psi^\perp} (\hat A + \hat B) \ket{\psi} \right|^2.
\label{eq_maccone_pati_2}
\end{equation}

Relations~\eqref{eq_maccone_pati_1} and~\eqref{eq_maccone_pati_2} are stronger than those based on the product of uncertainties, since an appropriate choice of the orthogonal state on the r.h.s. can lead to saturation of the inequality. For a comprehensive review, see for instance Ref.~\cite{Dodonov2025}.

Let us now consider an arbitrary multipartite quantum system. For any given $k$-partition in $k$ subsystems $\{A_1,\dots,A_k\}$, an arbitrary state is described by an element   $\rho \in \mathcal{D}(\mathcal{H})$. Here $\mathcal{D}(\mathcal{H})$ denotes the set of density operators acting on the global Hilbert space $\mathcal{H} = \bigotimes_{l=1}^{k} \mathcal{H}^{(A_l)}$, i.e. the tensor product of the local Hilbert spaces $\mathcal H^{(A_l)}$, each of finite dimension $d_{A_l}$,
so that $\mathcal H$ has finite dimension $d=\prod_{l} d_{A_{l}}$. 

For each subsystem $A_l$, let $D_{A_{l}} = d_{A_{l}}^{2}-1$ and denote by $\{ \sigma_{\alpha_{l}}^{(l)}\}_{\alpha_{l}=0}^{D_{A_{l}}}$ the set of generators of $\mathrm{SU}(d_{A_{l}})$ in the fundamental representation, along with the identity, i.e., the $D_{A_l}$ Hermitian matrices of dimension $d_{A_{l}} \times d_{A_{l}}$ acting on $\mathcal{H}^{(A_{l})}$ plus the identity matrix $\sigma_{0}^{(l)}$. We can then introduce the local collective observables $\sigma_{\alpha_{1}\dots\alpha_{k}} \equiv \sigma_{\alpha_{1}}^{(1)} \otimes \dots \otimes \sigma_{\alpha_{k}}^{(k)}$, their uncertainties $\Delta^{2}{\sigma_{\alpha_{1}\dots
\alpha_{k}}} \equiv \braket{\sigma_{\alpha_{1}\dots\alpha_{k}}^{2}} -  \braket{\sigma_{\alpha_{1}\dots\alpha_{k}}}^{2}$, and the total uncertainty $\mathcal{U}_{T}$:

%Let us denote by $\{ \sigma_{\alpha_{l}}^{(l)}\}_{\alpha_{l}=0}^{D_{A_{l}}}$ the set of the matrix representations of the generators of the $\mathrm{SU}(d_{A_{l}})$ algebras, with $l=1,\dots,k$, i.e. the Hermitian matrices of dimension $d_{A_{l}} \times d_{A_{l}}$ acting on $\mathcal{H}^{(A_{l})}$ plus the identity matrix $\sigma_{0}^{(l)}$, with $D_{A_{l}} = d_{A_{l}}^{2}-1$. We can then introduce the local collective observables $\sigma_{\alpha_{1},\dots,\alpha_{k}} \equiv \sigma_{\alpha_{1}}^{(1)} \otimes \dots \otimes \sigma_{\alpha_{k}}^{(k)}$, their uncertainties $\Delta^{2}{\sigma_{\alpha_{1},\dots,\alpha_{k}}} \equiv \braket{\sigma_{\alpha_{1},\dots,\alpha_{k}}^{2}} -  \braket{\sigma_{\alpha_{1},\dots,\alpha_{k}}}^{2}$, and the total uncertainty $\mathcal{U}_{T}$:

\begin{equation}
\label{total_uncertianty}
\mathcal{U}_{T} \, \, \equiv \sum_{i_1=1}^{D_{A_1}} \dots \sum_{i_k=1}^{D_{A_k}} \Delta^{2}{\sigma_{i_{1}\dots i_{k}}}\equiv 
\sum_{\text{comb}} \Delta^{2}{\sigma_{i_{1} \dots i_{k}}} \, .
\end{equation}
where $\sum_{\text{comb}}$ denotes the sum over all possible tensor products of the $\mathrm{SU}(d_{A_l})$ generators.

Here we will show that the above total uncertainty Eq.~\eqref{total_uncertianty} satisfies an \textit{exact equality} expressed exclusively in terms of the purities of the global and of the reduces states, as proven in Section~\ref{sec_generalized_uncertainty}, Eqs.~\eqref{general_formula} and ~\eqref{formula_scarti_purezze}, of the present work. We find that the structure of the relation reflects the existence of a conservation law on the uncertainties of the ${\mathrm{SU}}(N)$ observables: once the purities of the global and the reduced states are given, any relative variation of the local uncertainties must leave the total uncertainty invariant.

For a multiparty system that includes at least one single-qubit subsystem, exploiting the equality holding for the total uncertainty allows us to determine the rest $\mathcal{R}$ that saturates the  Robertson-Schr\"odinger inequality Eq.~\eqref{equality_scrhrodinger_unciertainty} for any pair of non-commuting $\mathrm{SU}(2)$ observables. Specifically, we find that $\mathcal{R}$ equals the qubit linear entropy, which is trivially related to the purity and to the two-Rényi entropy; when the global state is pure, $\mathcal{R}$ quantifies the bipartite entanglement between the qubit and the remainder of the system. Therefore, the missing term that saturates the Robertson-Schr\"odinger uncertainty relation enjoys a simple physical interpretation in terms of the correlations -- both classical and quantum -- between a single qubit and the environment. 

In fact, the relation between uncertainties and entanglement goes deeper. Bipartite and multipartite entanglement are fundamental forms of nonlocal quantum correlations, and their study is not only fundamental for improving our understanding of the foundations and structural aspects of quantum theory but also crucial in various applications, such as, e.g., quantum teleportation~\cite{Bennett1985}, metrology~\cite{Giovannetti2011}, cryptography~\cite{Ekert1991, Gisin2002, Pirandola2020}, many-body physics~\cite{Laflorencie2016, Illuminati2022, Illuminati2023} and more. Indeed, major progress in qualifying, quantifying, and characterizing entanglement has been achieved by relating bipartite and multipartite entanglement certification to the violation of uncertainty-based inequalities expressed in terms of correlation matrices, covariance matrices, and correlation matrices~\cite{Hofmann2003,Guhne2004,Guhne2007,Badziag2008,deVicente2011,Friis2019,Vitagliano2023}.

Here we will go some steps further by establishing for general multipartite systems a set of simple direct relations, as expressed by Eq.~\eqref{general_formula} and Eqs.~\eqref{formula_scarti_purezze}, \eqref{general_formula2}, between the total uncertainty equality Eq.~\eqref{total_uncertianty}, the global and reduced quantum state purities, and the squared Hilbert-Schmidt (HS) norm of the system $\mathrm{SU}(N)$ correlation matrix. Leveraging on these results, we will then derive a necessary condition for $k$-separability, a simple inequality enjoying two different, but equivalent, formulations, one expressed exclusively in terms of the purity of the global multipartite state and the purities of the reduced states, and one expressed exclusively in terms of the elements of the correlation matrix. Whenever the inequality is violated, the global state is guaranteed to possess some form of entanglement (bipartite and/or multipartite). Thanks to the structural properties of the criterion, we prove that for any finite-dimensional system the set of states that satisfy the $k$-separability inequality is convex, and we analyze the performance of the criterion both numerically and analytically, identifying the classes of states for which entanglement is most efficiently detected. The criterion becomes increasingly effective as the amount of entanglement grows. 

The fact that the HS norm of the $\mathrm{SU}(N)$ correlation matrix can be expressed entirely in terms of state purities (or two-Rényi entropies) has two further important consequences. The first one is that all previously known entanglement detection criteria valid in particular instances that are based on the HS norm of the correlation tensor can be straightforwardly reformulated in terms of purities, including the criteria for $k$-inseparability for states of $n$-qubit systems~\cite{Badziag2008} and for genuine multipartite entanglement of qudit systems introduced in Refs.~\cite{Badziag2008,deVicente2011,Friis2019}. 

The second important consequence is that quantum state purities can generally be measured efficiently, even via simple single-copy measurement schemes, as their reconstruction from experimental data does not require full quantum state tomography \cite{vanEnk2012,Elben2018,Yanay2021}. Moreover, the efficiency of these protocols does not degrade with increasing system size, as explicitly demonstrated, e.g., for $n$-qubit systems \cite{vanEnk2012}. Consequently, considering non-tomographic measurement schemes of quantum state purities, our formulation provides an experimentally friendly and scalable tool for entanglement detection and certification. Crucially, while the HS norm of the correlation tensor involves $\prod_{l=1}^{k} D_{A_l}$ terms, our approach requires a number of purities that scales only as $2^k - 1$, making its use particularly advantageous in the case of multipartite qudit systems with high local Hilbert space dimension.

For $n$-qubit states, and two dichotomic observables per party, we show that our inequality is also related to a necessary condition for Bell nonlocality~\cite{Zukowski2002}; we prove that the connection is nontrivial, in the sense that there are local states that do violate the inequality and hence are not $k$-separable. Such states include two-qubit Werner states, mixed four-qubit GHZ states, and various other classes of $n$-qubit states. In the elementary instance of two-qubit states we show that convexity leads to a very simple geometric interpretation of the criterion for Bell-diagonal states: the states that do not violate the inequality lie on the circumscribed sphere of the octahedron of separable Bell-diagonal states. For general two-qubit states we show that the Clauser-Horne-Shimony-Holt (CHSH) form of Bell nonlocality can be re-expressed in terms of the purities; this property allows to establish a simple hierarchical relation between the purity-based entanglement criterion and the CHSH condition for quantum nonlocality. 

The paper is organized as follows. In Section~\ref{sec_generalized_uncertainty}, we derive the exact uncertainty equality for $\mathrm{SU}(N)$ observables. Section~\ref{sec_saturation_sr_unicertainty} is devoted to computing the exact expression of the remainder $\mathcal{R}$ in the Robertson-Schr\"odinger uncertainty relation. In Section~\ref{sec_k_separability}, we apply these results to the study multipartite entanglement and we derive a necessary condition for $k$-separability. In Section~\ref{sec_exponential_advantage} we compare the scalings of the purity-based and the correlation-based expressions of the $k$-separability criterion and discuss the exponential advantage of the former over the latter. Finally, several illustrative examples and applications are studied in detail in Section~\ref{sec_examples}, while in Section~\ref{sec_conclusions} we review our results and discuss some possible outlooks. 

\section{Total uncertainty, quantum state purities, and $\mathbf{SU(N)}$ correlation matrices} \label{sec_generalized_uncertainty}

In this section, we derive the exact expression of the total uncertainty for $\mathrm{SU}(d_{A_{l}})$ observables with respect to a fixed $k$-partition of an $n$-partite quantum system. We show that this quantity depends linearly only on the purity of the global state of the system and on the purities of the reduced states of all possible subsystems $\{A_1,\dots,A_k\}$ for any given $k$-partition.

Let us first expand the global $k$-partite density operator $\rho$ in terms of the generators of the $\mathrm{SU}(d_{A_{l}})$ algebra:

\begin{equation}
\label{general_do}
    \rho = \frac{1}{d} \hspace{-0.3em} \sum_{\alpha_{1},\dots,\alpha_{k}=0} ^{D_{A_{1}} \dots D_{A_{k}}}  \hspace{-0.8em} t_{\alpha_{1} \dots \alpha_{k}} \sigma_{\alpha_{1} \dots \alpha_{k}}\, ,
\end{equation}
where $\sigma_{\alpha_{1} \dots \alpha_{k}} = \sigma_{\alpha_{1}}^{(1)} \otimes \dots \otimes \sigma_{\alpha_{k}}^{(k)}$, $d=\prod_{l} d_{A_{l}}$, and the generators of $\mathrm{SU}(d_{A_{l}})$ have been normalized such that $\text{Tr}\left[ \sigma_{\alpha}^{(l)}\sigma_{\beta}^{(l)} \right] = d_{A_{l}}\delta_{\alpha\beta}$, and $t_{\alpha_1\dots \alpha_k}\equiv\braket{\sigma_{\alpha_1\dots \alpha_k}}_\rho$ are the $k$-point correlation functions of $\mathrm{SU}(N)$ observables. The purity $\mathcal{P}^{(A_{1}\dots A_{k})}$ of state $\rho$ can then be expressed as follows:
\begin{equation}
\label{purity_nk}
    \mathcal{P}^{(A_{1}\dots A_{k})} = \frac{1}{d} \hspace{-0.2em} \sum_{\alpha_{1},\dots,\alpha_{k}=0} ^{D_{A_{1}},\dots,D_{A_{k}}}  \hspace{-0.8em} t_{\alpha_{1}\dots \alpha_{k}}^{2} = \frac{1}{d} \|T\|^{2}\, ,
\end{equation}
where $T$ is the correlation tensor of elements $t_{\alpha_{1}\dots \alpha_{k}}$, and $\|T\|^{2}$ is the squared HS norm of $T$.

For any set of $g$ subsystems of the $k$-partition, with $g \leq k$, and $i_{1} \neq \dots \neq i_{g} \in \{1,\dots, k \}$, consider the correlation matrices $t^{(A_{i_{1}}\dots A_{i_{g}})}$ with elements $t_{\alpha_{1}\dots \alpha_{k}}$, $\alpha_{l} = 0$ if $l \not\in \{i_{1},\dots,i_{g}\}$, and $\alpha_{l} = 1,\dots,D_{A_{l}}$ otherwise. Using Eq.~\eqref{purity_nk} and observing that
\begin{equation}
\label{useful_eq}
    \|T\|^{2} = 1 + \sum_{i} \|t^{(A_{i})}\|^{2} + \sum_{i,j}\|t^{(A_{i}A_{j})}\|^{2}+ \dots + \| t^{(A_{1}\dots A_{k})}\|^{2}\, , 
\end{equation}
the following equality holds:
\begin{equation}
\label{general_formula}
    \| t^{(A_{1}\dots A_{k})}\|^{2} = \sum_{g=1}^{k} \sum_{ \mathcal A\in \{ A\}_{g} } (-1)^{k-g} d_{ \mathcal A} \mathcal{P}^{\mathcal A} + (-1)^{k} \,,
\end{equation}
where, for a fixed $g \leq k$, $\mathcal A \equiv (A_{i_1}\dots A_{i_g}) \in \{A\}_g$, with $i_{1}<\dots<i_{g}$, is a subset of $g$ subsystems extracted from the original $k$-partition, and $\{A\}_g$ is the set of all such possible subsystems. Consequently, the sum over $\{A\}_{g}$ runs over all possible combinations (without repetitions) of $g$ subsystems of the $k$ parties:
\begin{equation}
    \sum_{\mathcal A \in  \{ A\}_{g} }d_{\mathcal A}\mathcal{P}^{\mathcal A}  \equiv \hspace{-1em} \sum_{i_{1} < i_{2} < \dots < i_{g} = 1}^{k} \hspace{-1.5em} d_{A_{i_{1}}} \dots d_{A_{i_{g}}} \mathcal{P}^{(A_{i_{1}}\dots A_{i_{g}})} \, .
\end{equation}
As an explanatory example of the notation, in Appendix~\ref{appendix_notation} we describe in detail the two-qubit case, while in Appendix~\ref{appendix_proof} we report the proof of Eq.~\eqref{general_formula}, obtained by induction.

In the following, the range of Greek tensor indices includes $0$, while the range of Latin tensor indices excludes it: $t^{(A_1\dots A_k)}$ is the tensor of components $t_{i_1\dots i_k}$.

Next, consider the mean-squared values of the $\mathrm{SU}(d_{A_l})$ generators
\begin{align}
    \Delta^{2}{\sigma_{i_{1}\dots i_{k}}} \equiv &\braket{\sigma_{i_{1}\dots i_{k}}^{2}}_{\rho} -  \braket{\sigma_{i_{1}\dots i_{k}}}_{\rho}^{2} \, .
\end{align}
Recalling the definition Eq.~\eqref{total_uncertianty} of the total uncertainty $\mathcal{U}_{T}$, and reminding that $\sigma_{\alpha_{1}\dots \alpha_{k}}^{2} = \mathds{1}_{d}$ and that $D_{A_{l}} = d_{A_{l}}^{2}-1$, we finally have
\begin{align}
\label{formula_scarti_purezze}
    \mathcal{U}_{T} = \mathcal{N} - \| t^{(A_{1}\dots A_{k})} \|^{2}\, ,
\end{align}
where $\| t^{(A_{1}\dots A_{k})} \|^{2}$ is given by Eq.~\eqref{general_formula} in terms of the quantum state purities, and 
\begin{equation}
    \mathcal{N} = \sum_{\text{comb}} 1 = \prod_{l=1}^{k} D_{A_{l}} \, ,
\end{equation}
where $\sum_{\text{comb}}$ denotes the sum over all possible tensor products of the $\mathrm{SU}(d_{A_l})$ generators. Finally, we can rewrite the HS norm of the correlation matrix $t$ in terms of the total uncertainty, as follows:
\begin{equation}
\| t^{(A_{1}\dots A_{k})} \|^{2} = \mathcal{N} - \mathcal{U}_{T} \, .
\label{general_formula2}
\end{equation}
  
Eq.~\eqref{general_formula} and Eq.~\eqref{general_formula2} hold for any quantum state of a composite system of finite Hilbert space dimension and for any fixed $k$-partition of the system. These relations are our two first fundamental results, as they allow to express the squared HS norm of the correlation matrices in terms of the purity of the global state of the system and the purities of the reduced states of the $k$ subsystems or, equivalently, in terms of the total uncertainty associated with all tensor products of the $\mathrm{SU}(d_{A_l})$ observables. Once all the purities are fixed, the total uncertainty is also fixed. States sharing the same set of purities can differ only in the distribution of the uncertainties among the individual terms appearing in Eq.~\eqref{formula_scarti_purezze}, while the total sum remains invariant. This expresses a conservation law for the total uncertainty.

\section{Saturation of the Robertson-Schr\"odinger uncertainty relation} \label{sec_saturation_sr_unicertainty}

\subsection{Systems comprising at least a single qubit}
Consider a composite quantum system partitioned in $k$ subsystems
$A_{1}, \dots, A_{k}$, where at least one subsystem, say $A_{l}$, is a
single qubit. For ease of notation, in the following we drop the subsystem index
and denote the qubit simply by $A\equiv A_{l}$.

Introducing $r_{i} \equiv t_{i}^{(A)}$ and recalling that $t_{i}^{(A)} = \braket{\sigma_{i}^{A}}$ we can rewrite~\eqref{general_formula} as follows:
\begin{align}
    &\sum_{i}r_{i}^{2} = 2\mathcal{P}^{(A)} - 1,  \\
    &\Delta{\sigma_{i,A}}^{2} \equiv \braket{ (\sigma_{i}^{A} - \braket{\sigma_{i}^{A}})^{2} } = 1 - r_{i}^{2}\, ,
\end{align}
so that
\begin{equation}
\label{delta_relation_subsystem}
    \sum_{i}\Delta^{2}{\sigma_{i,A}} = 2(2 - \mathcal{P}^{(A)})
\end{equation}
and
\begin{align}
    \Delta^{2}{\sigma_{i,A}} \Delta^{2}{\sigma_{j,A}} &= 1 - r_{i}^{2} - r_{j}^{2} + r_{i}^{2}r_{j}^{2} \notag \\
    &= r_{k}^{2} + r_{i}^{2}r_{j}^{2} + 2(1 - \mathcal{P}^{(A)})\, ,
    \label{H_R_generalized}
\end{align}
with $i \neq j \neq k$ labeling the three Pauli matrices.

By inspection we see immediately that Eq.\eqref{H_R_generalized} is indeed the Robertson-Schr\"odinger relation, Eq.\eqref{equality_scrhrodinger_unciertainty}
\begin{equation}
    \Delta^{2}{\sigma_{i,A}} \Delta^{2}{\sigma_{j,A}} = \braket{\sigma_{k}^{A}}^{2} + \braket{\sigma_{i}^{A}}^{2}\braket{\sigma_{j}^{A}}^{2} + 2(1-\mathcal{P}^{(A)})\, ,
\end{equation}
with the rest $\mathcal{R}$ being:
\begin{equation}
\mathcal{R} = 2(1 - \mathcal{P}^{(A)}) \, . 
\label{LinearEntropy}
\end{equation}
The above expression~\eqref{LinearEntropy} is in fact the linear entropy $S_L$ of the single-qubit state. If the total multipartite system is in a pure state, the rest $\mathcal{R} = S_L$ coincides, up to a factor, both with the negativity and the concurrence, and thus quantifies the bipartite entanglement between qubit $A$ and the remainder of the system. If the total system is in a mixed state, $\mathcal{R} = S_L$ quantifies the total amount of correlations (both classical and quantum) between qubit $A$ and the remainder of the system. These results establish a remarkable set of relations between uncertainty, entanglement, and correlations.

Considering systems comprising at least a single qutrit (or any other qudit of higher dimension), quantifying the product of two non-compatible observables is not as straightforward as in the case of Pauli operators. One can nevertheless still make use of Eqs.~\eqref{formula_scarti_purezze} and~\eqref{general_formula}, which allow to quantify the sum of the uncertainties of non-compatible $\mathrm{SU}(d_{l})$ observables in terms of the purity of the global state and the purities of the reduced states of its subsystems, for any $k$-partition.

\subsection{Systems comprising at least two qubits}
Consider next a $k$-partition $A_{1},\dots,A_{k}$, of a quantum system, with at least two subsystems $A_{l}\equiv A$ and $A_{m} \equiv B$ that are both single qubits, and take into account the class of $T$-diagonal states, defined as those two-qubit states such that $t_{ij}^{(A,B)} = 0$ for $i \neq j$, with $i,j = 1,2,3$. This class is relevant because any two-qubit state can be put in $T$-diagonal form through local unitary operations~\cite{Horodecki2009}, thus  preserving the state's purities, entanglement, and nonlocality.

Putting $t_{ij} \equiv t^{(A,B)}_{ij}$, we can rewrite Eq.~\eqref{general_formula} as
\begin{equation}
    \sum_{i=1}^{3} t_{ii}^{2} = 4 \mathcal{P}^{(AB)} - 2 \mathcal{P}^{(A)} - 2\mathcal{P}^{(B)} + 1\, .
\end{equation}
Considering the uncertainty
\begin{align}
    \Delta^{2}{\sigma_{ij}} &= \braket{(\sigma_{i}^{A} \otimes \sigma_{j}^{B} - \braket{\sigma_{i}^{A}\otimes \sigma_{j}^{B}})^{2}} \notag \\
    &= 1 - \braket{\sigma_{i}^{A} \otimes \sigma_{j}^{B}}^{2} = 1 - t_{ij}^{2}\, ,
\end{align}
we have
\begin{equation}
\label{delta_ii_relation_T_diag}
    \sum_{i=1}^{3} \Delta^{2}{\sigma_{ii}} = 2 ( \mathcal{P}^{(A)} + \mathcal{P}^{(B)} + 1 - 2\mathcal{P}^{(AB)} ) \, ,
\end{equation}
and, finally, the following expression of the Robertson-Schr\"odinger uncertainty:
\begin{align}
    \Delta^{2}{\sigma_{ii}} \Delta^{2}{\sigma_{jj}} &= \braket{\sigma_{kk}}^{2} + \braket{\sigma_{ii}}^{2}\braket{\sigma_{jj}}^{2} \notag \\
    &+ 2(\mathcal{P}^{(A)} + \mathcal{P}^{(B)} - 2\mathcal{P}^{(AB)} ) \, ,
\end{align}
where $\sigma_{ii} \equiv \sigma_{i} \otimes \sigma_{i}$. The rest $\mathcal{R}$ takes thus the form
\begin{equation}
\mathcal{R} = 2(\mathcal{P}^{(A)} + \mathcal{P}^{(B)} - 2\mathcal{P}^{(AB)} ) \, .
\label{RestTwoQubits}
\end{equation}
In the following, we will prove that the rest $\mathcal{R}$, expressed through Eq.~\eqref{RestTwoQubits} in terms of the global and local purities, enters directly a purity-based sufficient condition for entanglement and a purity-based necessary and sufficient condition for CHSH Bell nonlocality, as shown in Section~\ref{sec_examples}.

If we do not restrict to $T$-diagonal states, the generalization of Eq.~\eqref{delta_ii_relation_T_diag} is 
\begin{equation}
\label{delta_relation_total_system}
    \sum_{i,j=1}^{3}\Delta^{2}{\sigma_{ij}} = 2(\mathcal{P}^{(A)} + \mathcal{P}^{(B)} - 2\mathcal{P}^{(AB)} + 4)\, .
\end{equation}

\section{Conditions for $k$-separability and entanglement}
\label{sec_k_separability}

In this section, we will show how the expression~\eqref{general_formula} for the HS squared norm of the reduced correlation matrix $t^{(A_{1}\dots A_{k})}$ for a fixed $k$-partition can be used to derive a necessary condition for $k$-separability.

Despite the many significant results achieved so far in characterizing and quantifying bipartite entanglement~\cite{Bengtsson2020,Horodecki2009,PlenioVirmani2007}, determining in general whether a quantum state $\rho$ is entangled, and in what form, remains a formidable task in the multipartite case, as the structure of entanglement grows increasingly complex with the number of parties~\cite{Bengtsson2020,Gühne2009,Horodecki2009,Rudnicki2024}. On general grounds, the powerful PPT condition for bipartite entanglement~\cite{Peres1996,Horodecki1996,Simon2000,Adesso2007} is a particular instance of a general criterion, based on positive linear maps, that provides a necessary and sufficient condition for separability of multipartite quantum states, but is unfortunately in general non computable~\cite{Horodecki2001}.
Less general but quite useful and more easily computable criteria, both for bipartite and multipartite entanglement, include the Computable Cross Norm (CCN) criterion~\cite{Rudolph2000,Rudolph2003,Anwar2019}, the majorization criterion~\cite{Nielsen2001}, and the reduction criterion~\cite{Horodecki1999}.

Other approaches for entanglement detection are based on the violation of certain inequalities derived from uncertainty relations~\cite{Hofmann2003,Guhne2004} and correlation matrices~\cite{Guhne2007,Badziag2008}. In fact, the Hilbert-Schmidt norm of the correlation matrix $t$ plays a central role in entanglement criteria and entanglement witnesses, both bipartite and multipartite, as discussed on general grounds in Refs.~\cite{deVicente2011,Badziag2008,Friis2019}. Most of the above mentioned criteria have been reviewed in various works~\cite{Horodecki2009,Gühne2009,Vollbrecht2002,Friis2019}, while more recent developments and criteria for entanglement detection have been discussed in Refs.~\cite{klockl2015, Shang2018, Friis2019, Liu2023, Tavakoli2024, Rico2024, Liu2026}.

In the next subsection, exploiting the results derived in the previous sections, we will derive a purity-based necessary criterion for $k$-separability, while in Section~\ref{sec_exponential_advantage} we will show that this formulation yields an exponential reduction in the number of measurements required to determine whether a given state is entangled.

\subsection{Derivation of the criterion}
Let us consider an arbitrary $n$-composite quantum system described by a state $\rho \in \mathcal{D}{(\mathcal{H})}$, where $\mathcal{H} = \bigotimes_{i=1}^{n} \mathcal{H}^{(I_i)}$ is the Hilbert space of the composite system, and $\mathcal{D}(\mathcal{H})$ is the set of density operators acting on $\mathcal{H}$.
The state $\rho$ is fully ($n$-separable) iff there exists at least one set of states $\{ \rho_{i}^{(I_{l})} \in \mathcal{D}{(\mathcal{H}^{(I_{l})})}\:|\;l\in\{1,\dots,n\} \,,\,i=1,\dots,P\}$, for some integer $P$ and a probability distribution $\{ p_{i} \}_i$, such that

\begin{equation}
    \rho = \sum_{i} p_{i} \rho_{i}^{(I_{1})} \otimes \dots \otimes \rho_{i}^{(I_{n})}\:.
\end{equation}

\begin{figure}
    \centering
    \includegraphics[width=0.7\linewidth]{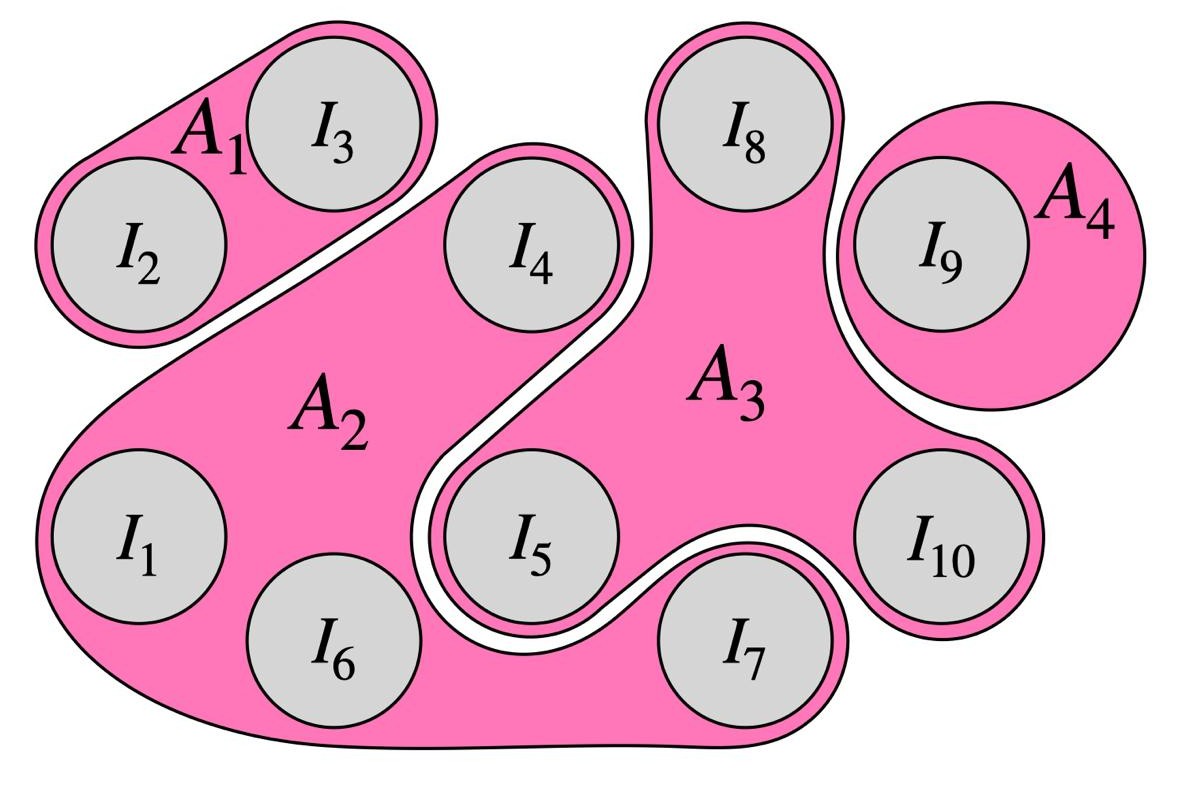}
    \caption{Example showing how $n$ factors entering in the full factorization may be combined to form a coarse-grained $k$-partition $k < n$. In this example $n=10$, $k=4$.}
    \label{fig:placeholder}
\end{figure}

If no such combination exists, the state is entangled in some form (bipartite and/or multipartite). If a state $\rho$ is not fully separable but only $k$-separable for a certain fixed $k$-partition of the system, with $k < n$ (see Fig.~\ref{fig:placeholder}), then $\rho$ admits the representation
\begin{equation}
\rho = \sum_{i} p_{i} \rho_{i}^{(A_{1})} \otimes \dots \otimes \rho_{i}^{(A_{k})}\, ,
\label{k-separability}
\end{equation}
with $\rho^{(A_{l})}  \in \mathcal{H}^{(A_{l})} \equiv \mathop{\bigotimes}_{s \in A_{l}} \hspace{-0.3em} \mathcal{H}^{(s)} $, for $l=1,\dots,k$ and $\{ A_{l} \}_{l=1}^{k}$ is a $k$-partition of the index set $I=\{ I_1,\dots,I_n \}$, that is $A_l \subset I$, $\bigcup_{l=1}^k A_l=I$, and $A_l\cap A_m=\emptyset$ if $l\neq m$. Clearly, if a state is fully separable, it is also $k$-separable for any $k$-partition, while the vice-versa in general does not hold. 

\subsubsection{Purity (entropy)-based formulation}
Consider the mean squared value $\braket{\Delta^{2}{O}}_{\rho}$ of a Hermitian operator $\hat{O}$ on a state $\rho$:
\begin{equation}
   \Delta^{2}_{\rho}{O} = \Big\langle \left( \hat{O} - \langle \hat{O} \rangle_{\rho} \right)^{2} \Big\rangle_{\rho}\, ,
\end{equation}
with $\braket{\hat{O}}_{\rho} \equiv \text{Tr}[\rho \hat{O}]$.

Expressing state $\rho$ as convex combination of states $\rho_{i}$ with probabilities $p_{i}$, we can write~\cite{Hofmann2003}:
\begin{equation}
\label{squared_mean_value_separable}
   \Delta^{2}_{\rho}{O} = \sum_{i}p_{i} \left( \Delta^{2}_{\rho_{i}}{O} + \widetilde{\Delta}_{\rho_{i}}^{2} \right)\, ,
\end{equation}
where $\widetilde{\Delta}_{\rho_{i}}^{2} = \left(\braket{\hat{O}}_{\rho_{i}} - \braket{\hat{O}}_{\rho}\right)^{2}$.

In principle, such relations allow to connect the separability problem to the violation of local uncertainty relations~\cite{Hofmann2003,Guhne2004,Guhne2007}. 

Although the sum of positive terms $\sum_{i} p_{i}\widetilde{\Delta}_{\rho_{i}}^{2}$ is in general not computable, we will now show how to evaluate it in the case of $\mathrm{SU}(N)$ observables.
Consider a $k$-separable state $\rho$ and observables that are tensor products of the Hermitian $\mathrm{SU}(d_{A_l})$ generators, with $l = 1, \dots, k$. Any operator $\rho^{(A_{l})}$ can be written as convex combination of the projectors corresponding to its (orthogonal) eigenvectors; therefore, without loss of generality, the $k$-separable state $\rho$, Eq.~\eqref{k-separability}, can be written as convex combination of pure $k$-product states:
\begin{equation}
    \rho = \sum_{i} \bar{p}_{i} \bigotimes_{l=1}^{k}\ket{\Psi_{i}}^{(A_{l})} {}^{(A_{l})}\bra{\Psi_{i}} 
\end{equation}
for some probability distribution $\{ \bar{p}_{i} \}$. 
%For a generic bipartite system, it has been shown that there exists a convex combination of $r$ product states with $r\leq d^{2}$~\cite{Horodecki1997}.
Inserting the $\mathrm{SU}(d_{A_k})$ observable $\hat{O} = \sigma_{i_{1}\dots i_{k}}$ in Eq.~\eqref{squared_mean_value_separable} and summing over $i_{1},\dots,i_{k}$, we have that the total uncertainty $\mathcal{U}_{T}$, Eq.~\eqref{total_uncertianty}, can be rewritten as follows:
\begin{align}
\label{Eq_scarti}
    \mathcal{U}_{T} =\sum_{j} \bar{p}_{j} \sum_{\text{comb}} \Delta^{2}_{\Psi_{j}} {\sigma_{i_{1}\dots i_{k}}} +  \widetilde{\Delta}^{2}\, ,
\end{align}
where $\sum_{\mathrm{comb}}$ is defined as in Eq.~\eqref{total_uncertianty},
\begin{equation}
    \widetilde{\Delta}^{2} = \sum_{j} \bar{p}_{j} \sum_{\text{comb}}\left(\braket{\sigma_{i_{1}\dots i_{k}}}_{\Psi_{j}} - \braket{\sigma_{i_{1}\dots i_{k}}}_{\rho}\right)^{2}
\end{equation}
is non-negative by definition, and the average $\braket{\dots}_{\Psi_{j}}$ is performed on the product state
\begin{equation}
    \ket{\Psi_{j}} \equiv \ket{\Psi_{j}}^{(A_{1})} \otimes \dots \otimes \ket{\Psi_{j}}^{(A_{k})}\, .
    \label{eq_psi_i}
\end{equation}
Eq.~\eqref{Eq_scarti} is equivalent to 
\begin{equation}
\widetilde{\Delta}^2=\mathcal{U}_T-\sum_j \bar{p}_j\mathcal{U}_{T,\Psi_j}\, ,
\end{equation} 
where $\mathcal{U}_{T,\Psi_j}$ is the total uncertainty Eq.~\eqref{total_uncertianty} computed on the product state $\ket{\Psi_{j}}$, Eq.~\eqref{eq_psi_i}. The total uncertainties $\mathcal{U}_T$ and $\mathcal{U}_{T,\Psi_j}$ can be computed using Eqs.~\eqref{general_formula} and~\eqref{formula_scarti_purezze}, with all the purities set to 1 for the pure product states $\ket{\Psi_j}$. Recalling that for a fixed $g \leq k$, $\mathcal A \equiv (A_{i_1}\dots A_{i_g}) \in \{A\}_g$, with $i_{1}<\dots<i_{g}$, is a subset of $g$ subsystems extracted from the original $k$-partition, and $\{A\}_g$ is the set of all such possible subsystems, we obtain for any $k$-separable state the following identity:
\begin{equation}
  \widetilde{\Delta}^{2} = \sum_{g=1}^{k} \sum_{\mathcal A\in\{A\}_g} (-1)^{k-g} d_{\mathcal A} \left(1 - \mathcal{P}^{\mathcal A} \right)\, .
  \label{delta_tilde}
\end{equation}
Since $\widetilde{\Delta}^{2}$ is non-negative, the r.h.s. of Eq.~\eqref{delta_tilde} must be non-negative for $k$-separable states. Therefore
\begin{equation}
    \sum_{g=1}^{k} \sum_{\mathcal A\in\{A\}_g} (-1)^{k-g} d_{\mathcal A} \left(1 - \mathcal{P}^{\mathcal A} \right) \geq 0
    \label{necessary_for_k_separab}
\end{equation}
provides a necessary condition for $k$-separability. Then, for a generic quantum state $\rho$ and a partition $\{ A_{l} \}_{l=1}^{k}$,
\begin{equation}
\label{general_inequality_criterion}
    \sum_{g=1}^{k} \sum_{ \mathcal A \in \{ A\}_{g} } (-1)^{k-g} d_{ \mathcal A} \left(1 - \mathcal{P}^{\mathcal A} \right) < 0\, 
\end{equation}
provides a sufficient condition for entanglement.

Recalling the relation between the state purity and the $2$-Rényi entropy $S_{2}[\rho] = - \log\left( \mathcal{P}\right)$, inequality~\eqref{general_inequality_criterion} can be recast in the entropic form
\begin{equation}
    \sum_{g=1}^{k} \sum_{ \mathcal A \in \{ A\}_{g} } (-1)^{k-g} d_{ \mathcal A} \left(1 - e^{-S_{2}\left[\rho^{(\mathcal A)}\right]}\right) < 0\, .
    \label{conditionentropic}
\end{equation}

\subsubsection{Direct correlation-matrix-based formulation}
Now, recalling Eq.~\eqref{general_formula} and denoting the correlation matrix $t^{(A_{1}\dots A_{k})} \equiv t$, by direct comparison we see that inequality~\eqref{general_inequality_criterion} can be rewritten as follows:
\begin{equation}
\label{general_inequality_criterion2_brutta}
    \|t\|^{2}>(-1)^k+\sum_{g=1}^k\sum_{\mathcal A\in\{A\}_g}(-1)^{k-g}d_{\mathcal A}\:.
\end{equation}

Inequality~\eqref{general_inequality_criterion2_brutta} can be further simplified, as its right-hand side can be rewritten as
\begin{equation} (-1)^k+\sum_{g=1}^k\sum_{\mathcal A\in\{A\}_g}(-1)^{k-g}d_{\mathcal A} = \prod_{g=1}^{k}\left(d_{A_{g} }-1 \right)\, , \nonumber
\end{equation}
and therefore:
\begin{equation}
     \|t\|^{2}> \prod_{g=1}^{k}\left(d_{A_{g} }-1 \right)\, .
     \label{general_inequality_criterion2}
\end{equation}

Eq.~\eqref{general_inequality_criterion} (or~\eqref{conditionentropic}) and Eq.~\eqref{general_inequality_criterion2} are two different but equivalent versions of the same sufficient criterion for entanglement, one expressed in terms of the state purities (entropies), and one expressed in terms of the correlation matrix elements. In Section~\ref{sec_exponential_advantage} we will compare their relative merits and advantages.

\subsubsection{$n$-qubit systems}
The criterion~\eqref{general_inequality_criterion} and its equivalent form~\eqref{general_inequality_criterion2} are valid for any quantum system of arbitrary dimension and any $k$-partition. For $n$-qubit systems $d_{A_{g}} = 2$, so that the necessary condition for $k$-separability and the sufficient condition for entanglement read, respectively, 
\begin{eqnarray}
&& \|t\|^{2} \leq 1 \, , 
\label{ineq_nqubit} \\
&& \nonumber \\
&& \|t\|^{2} - 1 > 0 \, .
\label{ineq_nqubit_suff}
\end{eqnarray}

Condition~\eqref{ineq_nqubit} is a weaker instance of a more general necessary condition for $k$-separability that reads
\begin{equation}
    \| t \|^{2} \leq t_{\mathrm{max}}\, ,
    \label{stronger_HS_criterion}
\end{equation}
where $t_{\mathrm{max}} \leq 1$ is the largest generalized Schmidt coefficient of the correlation matrix $t$~\cite{Badziag2008}. As it is seldom possible to determine $t_{\mathrm{max}}$, inequality~\eqref{ineq_nqubit} proves useful in testing the $k$-separability of $n$-qubit states.

For $n=2$, as any two-qubit state $\rho$ is locally unitarily equivalent to a $T$-diagonal state $\rho_{\mathrm{TD}}$~\cite{Horodecki2009}, that is the two-qubit state whose correlation matrix $t$ is diagonal, i.e. $t_{ij} = 0$ for $i \neq j$, the two states have the same entanglement. By using Eqs.~\eqref{general_formula} and~\eqref{RestTwoQubits}, inequality~\eqref{ineq_nqubit} can thus be rewritten as
\begin{equation}
    \mathcal{R} < 0 \, ,
\end{equation}
where $\mathcal{R}$, Eq.~\eqref{RestTwoQubits}, is the rest in the Robertson-Schr\"odinger relation, evaluated for the $T$-diagonal state $\rho_{\mathrm{TD}}$ corresponding to the original two-qubit state $\rho$. This simple example shows once more the intimate and intricate relation between uncertainties, purities, entropies, correlation matrix elements, and the quantum state separability and entanglement properties.

\subsubsection{Convexity}
The set of states satisfying condition Eq.~\eqref{necessary_for_k_separab} is convex. Indeed, consider two states $\rho$ and $\sigma$ violating condition Eq.~\eqref{general_inequality_criterion2} and denote its r.h.s. by $X$. Then,
\begin{equation}
    \begin{split}
        &\|pt_\rho+(1-p)t_\sigma \|^{2}\\
        \leq&[p^2+(1-p)^2]X+2p(1-p)\braket{t_\rho,t_\sigma}_\text{HS}\\
         \leq & [p^2+(1-p)^2]X+2p(1-p)\sqrt {\|{t_\rho\|^{2}\:\|t_\sigma\|^{2}}}\\
         \leq & X\:,
    \end{split}
\end{equation}
where $\braket{A,B}_{\mathrm{HS}}$ is the HS scalar product between matrices $A$ and $B$, and in the derivation of the second inequality we applied the Cauchy-Schwarz inequality.

\subsection{Other $t$-HS-norm-based criteria and their formulation in terms of purities}
Criteria involving the Hilbert-Schmidt norm of the correlation matrix $t$ are not restricted to the $k$-separability problem but include also more general instances, such as conditions on genuine multipartite entanglement and Bell nonlocality. We will briefly review some of these criteria and how they can also be expressed and evaluated in terms of purity (entropy)-based formulations, thanks to the general relation Eq.~\eqref{general_formula}. In Section~\ref{sec_exponential_advantage}, we will discuss how the formulation based on purities (entropies) provides an exponential advantage for the detection of entanglement and nonlocality.

\subsubsection{Genuine multipartite entanglement}
For quantum systems of arbitrary dimensions, the violation of inequalities of the form
\begin{equation}
    \|t\|^{2} \leq f(d_{l}) 
\label{genuinemultipartite}
\end{equation}
implies genuine multipartite entanglement~\cite{deVicente2011}. In the above, $f(d_{l})$ is a positive function that depends only on the dimensions of the system and its subsystems. For instance, in the case of three qudits $A_{1},A_{2},A_{3}$, with the same local dimension $d_{l}$, one has
\begin{equation}
\label{multipartite_criterion_3_parts}
    \|t^{(A_{1}A_{2}A_{3})}\|^2 \leq \frac{8 (d_{l}-1)(d_{l}^{2}-1)}{d_{l}^{3}}\, .
\end{equation}

Clearly, we can use Eq.~\eqref{general_formula} to rewrite the l.h.s. of Eq.~\eqref{genuinemultipartite} in terms of the purities (or the two-Rényi entropies) of the global and reduced states, so that the sufficient condition for genuine multipartite entanglement becomes
\begin{equation}
    \sum_{g=1}^{k} \sum_{ \mathcal A\in \{ A\}_{g} } (-1)^{k-g} d_{ \mathcal A} \mathcal{P}^{\mathcal A} + (-1)^{k} \leq f(d_{l})\, ,
    \label{other_criterion}
\end{equation}
and in the particular case of three qudits, condition Eq.~\eqref{multipartite_criterion_3_parts} becomes
\begin{equation}
\begin{split}
d_{l}^3 \mathcal{P}^{(A_{1}A_{2}A_{3})} + d_{l} \sum_{i=1}^{3} \mathcal{P}^{(A_{i})}
&- d_{l}^2 \sum_{i,j=1}^{3} \mathcal{P}^{(A_{i}A_{j})} - 1 \\
&\leq \frac{8 (d_{l}-1)(d_{l}^2-1)}{d_{l}^3} \,.
\end{split}
\end{equation}

\subsubsection{Relation to Bell inequalities for $n$-qubit states}
Consider an $n$-qubit system in a scenario involving two dichotomic measurements per party. It has been shown that if the state is Bell nonlocal, there exists a set of local coordinates such that~\cite{Zukowski2002}
\begin{equation}
\sum_{i_{1},\dots,i_{n}=1}^{2} t_{ij}^{2} > 1\, .
\label{inequality_nonloc}
\end{equation}
Inequality~\eqref{inequality_nonloc} is thus a necessary condition for Bell nonlocality of $n$-qubit states. Observing that the summation runs over only two of the three indices of the correlation matrix $t$ for each qubit, it follows that 
inequality~\eqref{inequality_nonloc} implies inequality~\eqref{ineq_nqubit}, which, in contrast, is invariant under local coordinate changes. 
On the other hand, condition~\eqref{inequality_nonloc} is not sufficient for Bell nonlocality, since there exist local states that violate it, so that the separability condition~\eqref{ineq_nqubit} can detect Werner-like states admitting a local hidden-variable description as we will show with some specific examples in Section~\ref{sec_examples}.

\section{Exponential advantage of the formulation in terms of purities}
\label{sec_exponential_advantage}
Exploiting the relation Eq.~\eqref{general_formula} connecting the HS squared norm of the correlation matrix $t$ to the state global and reduced purities, we have shown that a necessary condition for $k$-separability -- sufficient for entanglement -- can be written in two different but equivalent forms, Eqs.~\eqref{general_inequality_criterion} and~\eqref{general_inequality_criterion2}, and that in fact, again thanks to Eq.~\eqref{general_formula}, any criterion involving the norm of $t$~\cite{Badziag2008,deVicente2011,Friis2019}, such as Eq.~\eqref{other_criterion} for genuine multipartite entanglement and Eq.~\eqref{inequality_nonloc} for Bell nonlocality, can be reformulated in terms of state purities (or two-Rényi entropies). 

In this Section we will compare the two different formulations, respectively in terms of correlation matrix elements and in terms of state purities, and we will show that the latter brings about a significant advantage in actual experimental situations. The key reason for this advantage is that the number of quantum state purities to be measured is exponentially smaller than the number of the correlation matrix elements. Moreover, purities can be efficiently measured using techniques such as randomized measurement schemes~\cite{vanEnk2012,Elben2018,Yanay2021}. Importantly, the efficiency of such schemes improves as the number of parties increases, as demonstrated, for instance, in the $n$-qubit case~\cite{vanEnk2012}.

To begin with, we will first investigate the different scaling properties of $t$-matrix elements and state purities in the concrete example of a generic, unknown $n$-qubit state $\rho$ for a fixed $k$-partition. For any unknown $\rho$, taking into account the Pauli observables, one needs to measure $3^{k}$ terms to determine $\|t\|^{2}$, and the situation becomes exponentially more demanding for higher-dimensional qudit systems.
In contrast, for any fixed $k$-partition, the number of independent parameters required to determine the purities is only $2^{k}-1$. The relative scalings are reported in Table~\ref{tab:correlation_vs_purities}.

\begin{table}[h]
\centering
\begin{tabular}{c c c}
\hline\hline
Local dimension & $\#$ of matrix elements & $\#$ of purities \\
\hline
Qubits $(d_i=2)$ & $3^{k}$ & $2^{k}-1$ \\
Qudits $(\text{uniform:} \, \, d_i=d_1)$ & $(d_1^{2}-1)^{k}$ & $2^{k}-1$ \\
Qudits $(d_i\ \text{arbitrary})$ & $\displaystyle \prod_{i=1}^{k} (d_i^{2}-1)$ & $2^{k}-1$ \\
\hline\hline
\end{tabular}
\caption{Comparison between the number of correlation-tensor components required to evaluate $\|t\|^{2}$ and the number of purities for a fixed $k$-partition.}
\label{tab:correlation_vs_purities}
\end{table}

Although the time required to experimentally estimate a single purity element, $\tau_{\mathrm{pur}}$, may differ from the time needed to measure a single element of the correlation matrix, $\tau_{\mathrm{corr}}$, it has been shown that $\tau_{\mathrm{pur}}$, defined as the number of measurements required to estimate a single purity element within a fixed error, does not increase with the size of the system (and thus with the number of elements of the correlation matrix), as proved, e.g., in Ref.~\cite{vanEnk2012}.  
Therefore, the total measurement times for the two formulations are, respectively:
\begin{align}
&\mathcal{T}_{\mathrm{corr}} = (\# \text{ correlation matrix elements}) \cdot \tau_{\mathrm{corr}}, \notag \\
&\mathcal{T}_{\mathrm{pur}} = (\# \text{ purities}) \cdot \tau_{\mathrm{pur}}.
\end{align}
Since the above quantities obey the scaling behaviors reported in Table~\ref{tab:correlation_vs_purities}, it follows that for sufficiently large system sizes $\mathcal{T}_{\mathrm{pur}} \ll \mathcal{T}_{\mathrm{corr}}$, even when $\tau_{\mathrm{corr}} < \tau_{\mathrm{pur}}$.

The origin of the scaling is clear: the Hilbert-Schmidt norm of the correlation matrix, as well as the global and local purities, are invariant under local unitary transformations. Therefore, our formula identifies precisely the minimal set of features necessary to determine $\|t\|^{2}$.

It is true that, as noted in Ref.~\cite{deVicente2011}, it is not necessary to measure all the correlation terms, since each of them contributes a positive quantity. Hence, once the partial sum exceeds $1$, it is already guaranteed that $\|t\|^{2} > 1$. Nevertheless, for any state in $\mathcal{D}(\mathcal{H})$ this situation does not typically occur. 
This observation can be understood using the following argument. A mixed state $\rho$ can be sampled according to the probability distribution $P(\rho)= P_H(\textbf{F}^{(d)})\times P(\vec \lambda)$, where $P_H(\textbf{F}^{(d)})$ is the distribution of the flag manifold induced by the Haar measure, and $P(\vec \lambda)$ is a distribution on the eigenvalues simplex, which can be chosen in different ways~\cite{Zyczkowski2001}. 
For large $d$, different possible relevant choices, such as the Hilbert-Schmidt and Bures distributions, all lead to a behavior of the average purity $\langle \mathcal P^{(A_{1}\dots A_{k})} \rangle$ of the form $\langle \mathcal P^{(A_{1}\dots A_{k})} \rangle \sim \frac{\alpha}{d}$, with $\alpha$ a parameter of order unity~\footnote{Specifically, we have $\langle \mathcal P^{(A_{1}\dots A_{k})} \rangle_\mathrm{HS} = {2d}/(d^2+1)$ and $\langle \mathcal P^{(A_{1}\dots A_{k})} \rangle_\mathrm{B} = (5d^2+1)/(2d(d^2+1))$, corresponding to $\alpha=2$ and $\alpha=5/2$, respectively.}~\cite{Zyczkowski2001,Bengtsson2020}. Therefore, by using Eq.~\eqref{purity_nk}, one obtains $\|T\|^{2} \sim \alpha$, despite $d$ being large, and the number of correlation terms required to determine whether $\|t\|^{2} > 1$ scales as reported in Table~\ref{tab:correlation_vs_purities}.

\begin{figure}[h]
    \centering
    % Prima immagine come sottofigura
    \begin{subfigure}{0.35\textwidth}
        \centering
        \includegraphics[width=\linewidth]{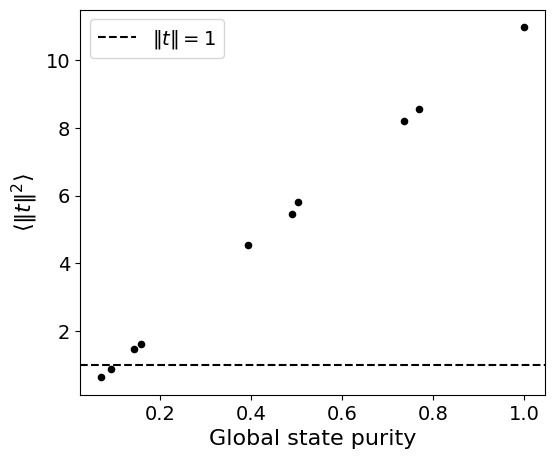}
        \caption{Mean value $\langle \|t\|^{2} \rangle$ of $\| t\|^{2}$ as a function of the global state purity, averaged over $50$ randomly extracted states per each fixed value of the purity.}
        \label{Fig:t_averaged}
    \end{subfigure}
    \hfill
    % Seconda immagine come sottofigura
    \begin{subfigure}{0.35\textwidth}
        \centering
        \includegraphics[width=\linewidth]{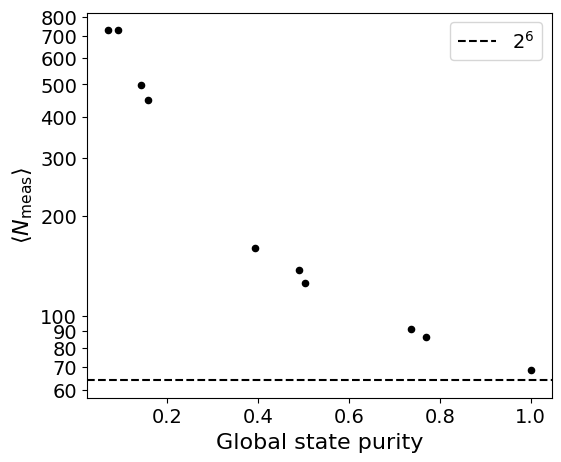}
        \caption{Average minimal number $\langle N_{\mathrm{meas}} \rangle$ of $t_{ij}$ elements required to determine if $\|t\|^{2} > 1$ as a function of the global state purity, computed over $50$ randomly extracted states per each fixed value of the purity. When $\|t\|^{2}<1$, this number reaches $3^{6}=729$, while the number of purities needed to compute $\|t\|^{2}$ exactly is $2^{6}-1=63$. The ordinate axis is in logarithmic scale.}
        \label{Fig:n_averaged}
    \end{subfigure}
    \caption{Example of the exponential advantage of the purity-based formulation for a system of $n = 6$ qubits with $k = n$.}
    \label{Fig:combined_figure_2}
\end{figure}

In order to better understand the regimes in which the purity-based formulations provide an exponential advantage, let us consider for example the case of $n = 6$ qubits with $k = n$. In Fig.~\ref{Fig:t_averaged}, we plot the average value $\langle \|t\|^{2} \rangle$ of $\|t\|^{2}$ as a function of the global state purity, averaged over 50 randomly extracted states for each fixed value of the purity. In Fig.~\ref{Fig:n_averaged}, we report the average minimal number $\langle N_{\mathrm{meas}} \rangle$ of elements $t_{ij}$, randomly extracted from $t$, that are required to ensure that $\| t \|^{2} > 1$. 

The previous discussion and the examples show clearly that there is no advantage in using the purity relation to compute $\|t\|^{2}$ for pure states. However, as the purity decreases, $\|t\|^{2}$ also decreases, causing $\langle N_{\mathrm{meas}} \rangle$ to grow exponentially. This demonstrates that Eq.~\eqref{general_formula} provides an exponential advantage in such cases.

Considering the general case of qudit systems, the exponential advantage becomes extremely pronounced, as summarized in Table~\ref{tab:correlation_vs_purities}. Indeed, the number of elements of the correlation matrix grows ever faster with the system dimension, since the number of $\text{SU}(d)$ generators increases quadratically with $d$. On the other hand, irrespective of the dimension, the number of purities keeps being always $2^{k}$ for any fixed $k$-partition.
For instance, if we consider the criterion for genuine multipartite entanglement, inequality~\eqref{multipartite_criterion_3_parts}, in the minimal case of a tripartite ($k=3$) three-qudit system with uniform local dimension $d_i = d_1$, the number of elements of the correlation matrix scales as $(d_{1} - 1)^{6}$, which is $64$ for $d_1 = 3$, and already $729$ for $d_1 = 4$, while the number of purities is always only $2^{3} - 1 = 7$, irrespective of $d_{i}$.

Finally, if the state $\rho$ is known, in the bipartite case of two qudits with uniform local dimension $d_{A}$, the experimental effort required to measure $\|t\|^{2}$ can be reduced to $d_{A}^{2}$ terms by exploiting the Schmidt decomposition, and there is no advantage in resorting to the purity-based formulation. On the other hand, in the multipartite case such a simplification is not possible, since a general Schmidt decomposition does not exist for $k$ parties. Consequently, the purity-based formulation retains its usefulness in the multipartite case, even when the state of the system is known.

\section{Examples} \label{sec_examples}
In this section we will study and discuss inequality~\eqref{general_inequality_criterion} for various particular cases.  

\subsection{2-qubit states}
We start from the simplest case of two qubits, $A$ and $B$, for which the HS squared norm of the correlation matrix is given by formula Eq.~\eqref{bipartite_qubits_formula} of Appendix \ref{appendix_notation}, so that the sufficient condition for entanglement (necessary for separability) takes the simplest form
\begin{equation}
\label{criterion_two_qubits_a}
\|t\|^2 - 1 \, = \, 4\mathcal{P}^{(AB)} - 2(\mathcal{P}^{(A)} + \mathcal{P}^{(B)}) \, >  \, 0 \, ,
\end{equation}

which, for T-diagonal states reduces to
\begin{equation}
\label{criterion_two_qubits_b}
\mathcal{R} \, <  \, 0 \, .
\end{equation}

The above relations cast the correlation matrix-based separability condition directly in the form of an inequality for the quantum state purities in general and, in the case of T-diagonal states, also for the rest $\mathcal{R}$ in the Robertson-Schrödinger uncertainty relation, showing explicitly and in the simplest possible way how a direct connection between correlations, purities (entropies), and uncertainties enters beautifully in the characterization of the state separability problem. In the case of pure states, $\mathcal{P}^{(AB)} = 1$, $\mathcal{P}^{(A)} = \mathcal{P}^{(B)} \leq 1$, the condition is necessary and sufficient, with $\mathcal{R} = 0$ holding iff $\rho$ is a product state.

\subsubsection{Relation to Bell nonlocality}

For two-qubit states, the rest $\mathcal{R}$ enters also in the characterization of Bell nonlocality as defined by the violation of the Clauser-Horne-Shimony-Holt (CHSH) inequality. Indeed, for a two-qubit state the necessary and sufficient condition for the violation of the CHSH inequality reads~\cite{Horodecki1995}
\begin{equation}
\label{CHSH_inequality}
    u_{1} + u_{2} > 1 \, ,
\end{equation}
where $u_{1}$ and $u_{2}$ are the two largest eigenvalues of the matrix $U = t^{T} t$ where $T$ denotes transposition. Denoting by $u_{3}$ the smallest eigenvalue of $U$, one has
\begin{equation}
    \|t\|^{2} = \sum_{i=1}^{3}u_{i} \, .
\end{equation}
Therefore, recalling again Eq.~\eqref{bipartite_qubits_formula}, we can rewrite the CHSH inequality~\eqref{CHSH_inequality} as follows:
\begin{equation}
\label{CHSH_condtion_purities}
4\mathcal{P}^{(AB)} -2(\mathcal{P}^{(A)} + \mathcal{P}^{(B)}) \, = \, - \, \mathcal{R} \, > \, u_{3} \, ,
\nonumber
\end{equation}
that is
\begin{equation}
\mathcal{R} \, < \, - \, u_{3} \, .
\end{equation}
As $u_{3} \geq 0$, if a T-diagonal two-qubit state violates the CHSH inequality, then it violates also the inequality~\eqref{criterion_two_qubits_b}. 

\subsubsection{Werner states and Bell-diagonal states - a geometrical interpretation}
\begin{figure}[t!]
    \centering{\includegraphics[width=.45\textwidth]{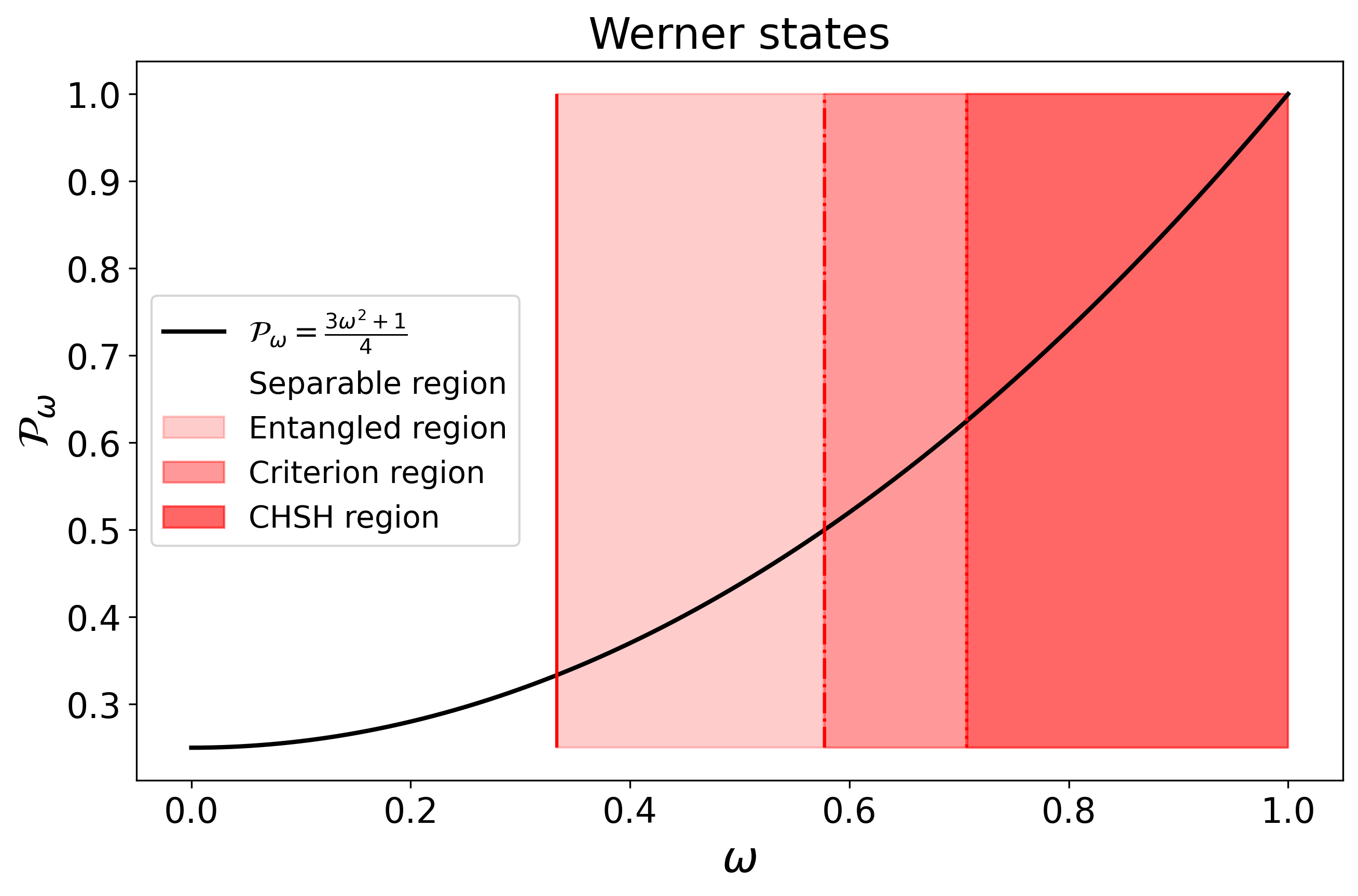}}
    \caption{Purity of two-qubit Werner states $\mathcal{P}_{\omega}$ as a function of $\omega$, spanning the different regions where the states are: maximally mixed, $\omega = \frac{1}{4}$; separable, $\omega \leq \frac{1}{3}$; entangled, $\omega > \frac{1}{3}$; in which the sufficient criterion is able to detect entanglement, $\omega > \frac{1}{\sqrt{3}}$; and, finally, where the CHSH inequalities are violated, $\omega > \frac{1}{\sqrt{2}}$. The threshold values of the purities at the boundaries between the various regions are, respectively: $\mathcal{P}_\mathrm{sep} = \frac{1}{3}$; $\mathcal{P}_\mathrm{criterion} = \frac{1}{2}$; and $\mathcal{P}_\mathrm{CHSH} = \frac{5}{8}$.}
    \label{Fig:purity_Werner}
\end{figure}

Given the above results, we wish to understand what relations hold between uncertainties, entropies, and correlations for entangled two-qubit states that do not violate the CHSH inequality. To this end, consider the Werner states~\cite{Werner1989}, i.e. the convex combinations of a Bell state with the maximally mixed state:
\begin{equation}
    \rho_{\omega} = \omega \ket{\Psi}\bra{\Psi} + \frac{1-\omega}{4} \mathds{1}_{4}\, ,
\end{equation}
where $\ket{\Psi}$ is any one of the four Bell states, $\mathds{1}_{4}$ is the identity operator in the four-dimensional Hilbert space of the two qubits, and $ 0 \leq \omega \leq 1$ is the convex superposition parameter. Werner states are entangled for $\omega > \frac{1}{3}$ and violate the CHSH inequality for $\omega > \frac{1}{\sqrt{2}}$. The reduced states of a Werner state are maximally mixed -- $\mathcal{P}^{(A)} = \mathcal{P}^{(B)} = \frac{1}{2}$ -- so that inequality~\eqref{criterion_two_qubits_a} is violated for $\omega > \frac{1}{\sqrt{3}}$, as already noticed in Ref.~\cite{Guhne2004}. Although the condition is only sufficient, it still detects entanglement in states that do not violate the CHSH inequality, as graphically summarized in Fig.~\ref{Fig:purity_Werner}.

Werner states are a particular case of the class of Bell-diagonal (BD) states, i.e. states that are convex combinations of Bell states. BD states admit the following parameterization~\cite{Horodecki1995}:
\begin{equation}
    \rho_\mathrm{BD} = \frac{1}{4} \left(1 + \sum_{i=1}^{3} t_{ii} \sigma_{i}^{A} \otimes \sigma_{i}^{B} \right)\, ,
\end{equation}
where the correlation matrix $t$ is diagonal: $t_{ij} = 0$ for $i \neq j$, and Werner states are obtained for $t_{ii} = -\omega$ for $i = 1,2,3$.
%By using Eq.~\eqref{bipartite_qubits_formula}, for two-qubit states, the inequality~\eqref{criterion_two_qubits} takes %the form
%\begin{equation}
%\label{geometric_two_qubits}
%    \| t^{(AB)} \|^{2} - 1 > 0\, .
%\end{equation}

In general, given two arbitrary quantum states $\rho_1$, $\rho_2$ of finite dimension $d$, and their associated correlation tensors $T_1$, $T_2$, it is easy to verify that the HS distance between the states coincides with the HS distance between the correlation tensors: 
\begin{equation}
\| \rho_1 - \rho_2  \|^{2} \, = \, \frac{1}{d}\| T_1 - T_2 \|^{2} \, .
\label{HS_distance}    
\end{equation}
For BD states, it is straightforward to verify that $ \| T_{\mathrm{BD}} \|^{2} = 1 + \| t_{\mathrm{BD}} \|^{2}$, so that the HS distance in Hilbert space between BD states coincides with the HS distance in matrix space of the associated correlations matrices:
\begin{equation}
\| \rho_{\mathrm{BD}_1} - \rho_{\mathrm{BD}_2}  \|^{2} \, = \, \frac{1}{4}\| t_{\mathrm{BD}_1} - t_{\mathrm{BD}_2} \|^{2} \, .
\label{HS_distanceBD}    
\end{equation}
Recalling that the maximally mixed (MM) state is still a BD state and that its correlation matrix $t_{MM}$ is the null matrix (actually, this is true in any dimension: the correlation matrix of any MM state is always the null matrix), it follows that the squared HS norm of the correlation matrix $\| t_{\mathrm{BD}}\|^{2}$ is the 
HS distance of a BD state from the MM state. 

\begin{figure}
    \includegraphics[width=0.5\linewidth]{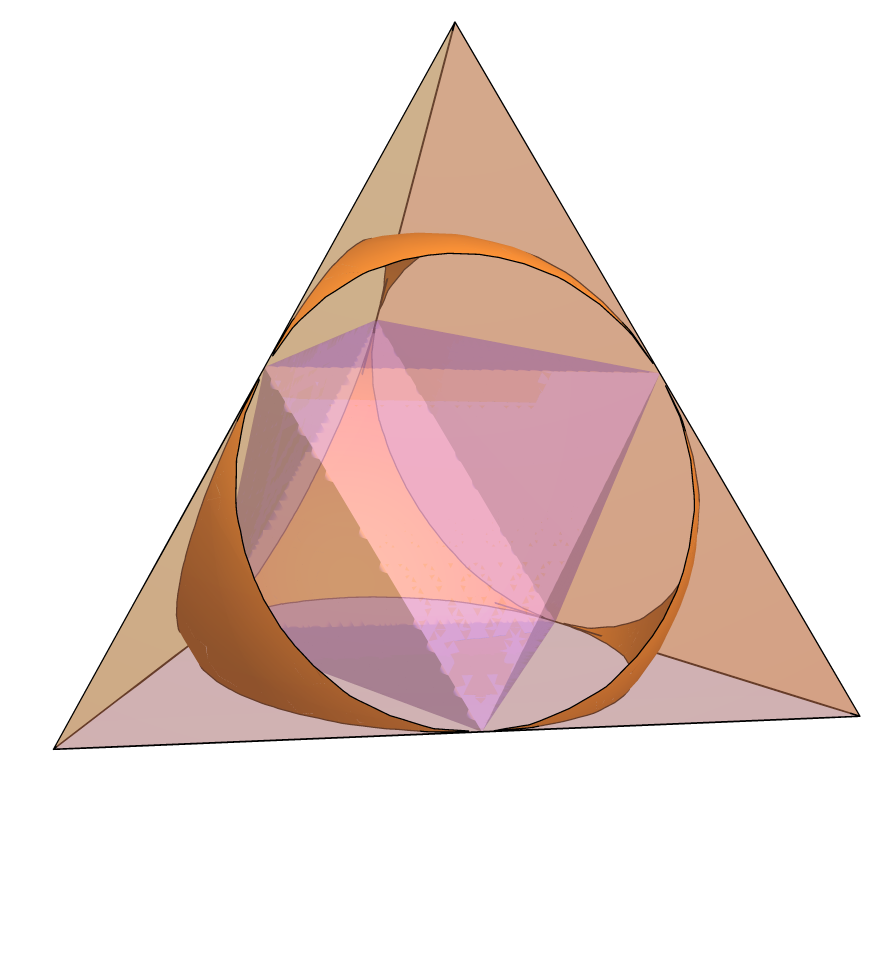}
    \caption{Tetrahedron of all the BD states, together with the octahedron of the BD separable states and the criterion demarcation surface. BD states lying outside of the surface are the entangled BD states detected by the criterion.}
    \label{fig:criterion_sphere_bd}
\end{figure}

In $t-$space, the criterion demarcation surface $\| t \|^{2} = 1$ identifies a sphere of BD states with constant purity $\mathcal{P}_{\mathrm{BD}}^{(AB)} = \frac{1}{2}$, while the set of separable BD states is the octahedron $|t_{11}| + |t_{22}| + |t_{33}| = 1$ inscribed in the sphere, as shown in Fig.~\ref{fig:criterion_sphere_bd}. 
From this geometry, one can estimate the fraction $R_{\mathrm{BD}}^{detected}$ of entangled BD states detected by the criterion with respect to the total number of entangled BD state by computing the ratio between the corresponding two volumes in $t-$space:
\begin{equation}
 R_{\mathrm{BD}}^{detected} = \frac{2\sqrt{3} - \pi \sqrt{3} + \pi}{2\sqrt{6}} \simeq 0.52 \, .
\end{equation}

\subsubsection{Randomly extracted two-qubits states}
\begin{figure}[h!]
    \centering
    % Prima immagine come sottofigura
    \begin{subfigure}{0.45\textwidth}
        \centering
        \includegraphics[width=\linewidth]{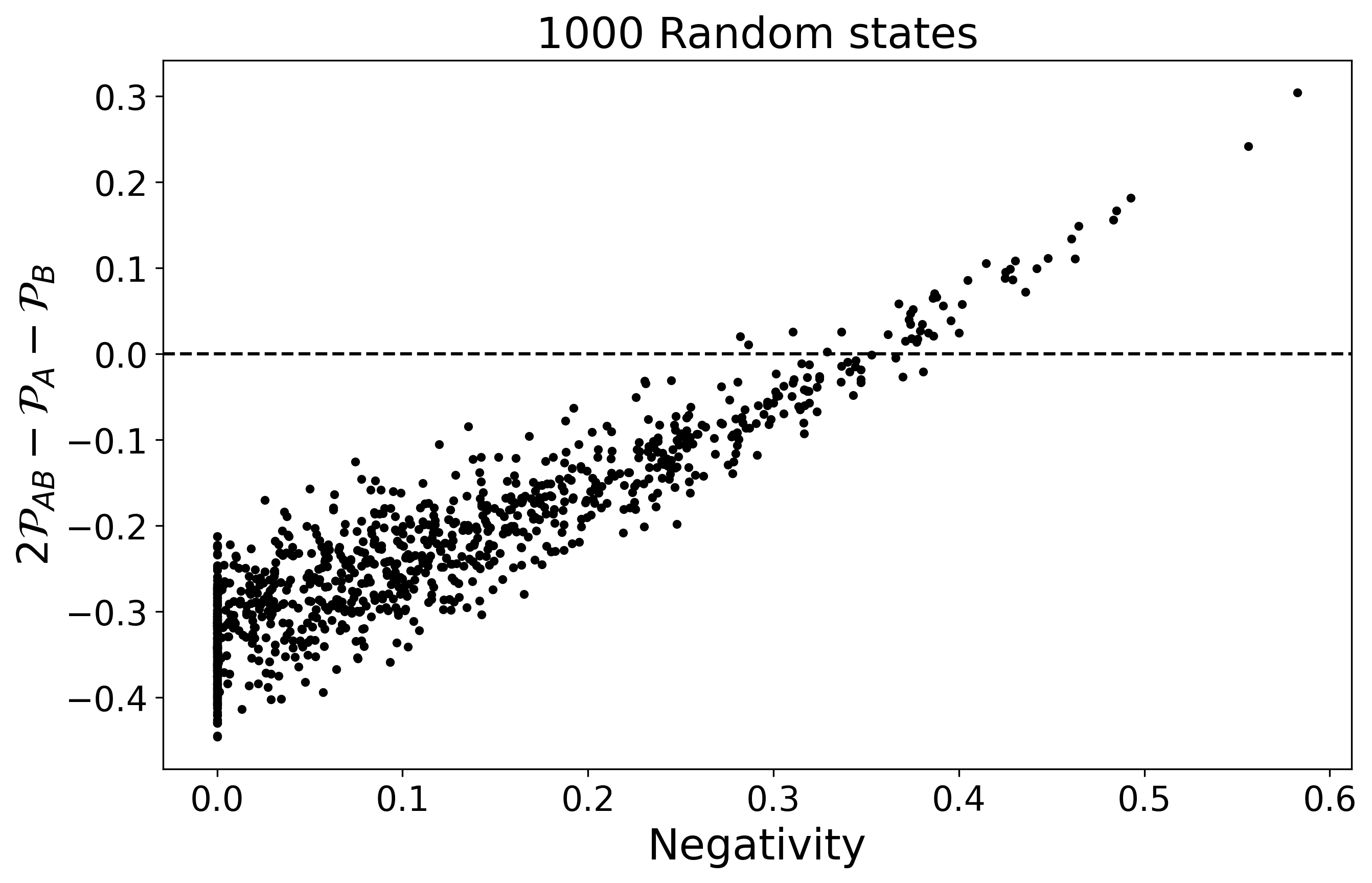}
        \caption{Generic random extracted two-qubit states; the criterion fails for values of the negativity $\leq 0.4$.}
        \label{Fig:pur_random_plot}
    \end{subfigure}
    \hfill
    % Seconda immagine come sottofigura
    \begin{subfigure}{0.45\textwidth}
        \centering
        \includegraphics[width=\linewidth]{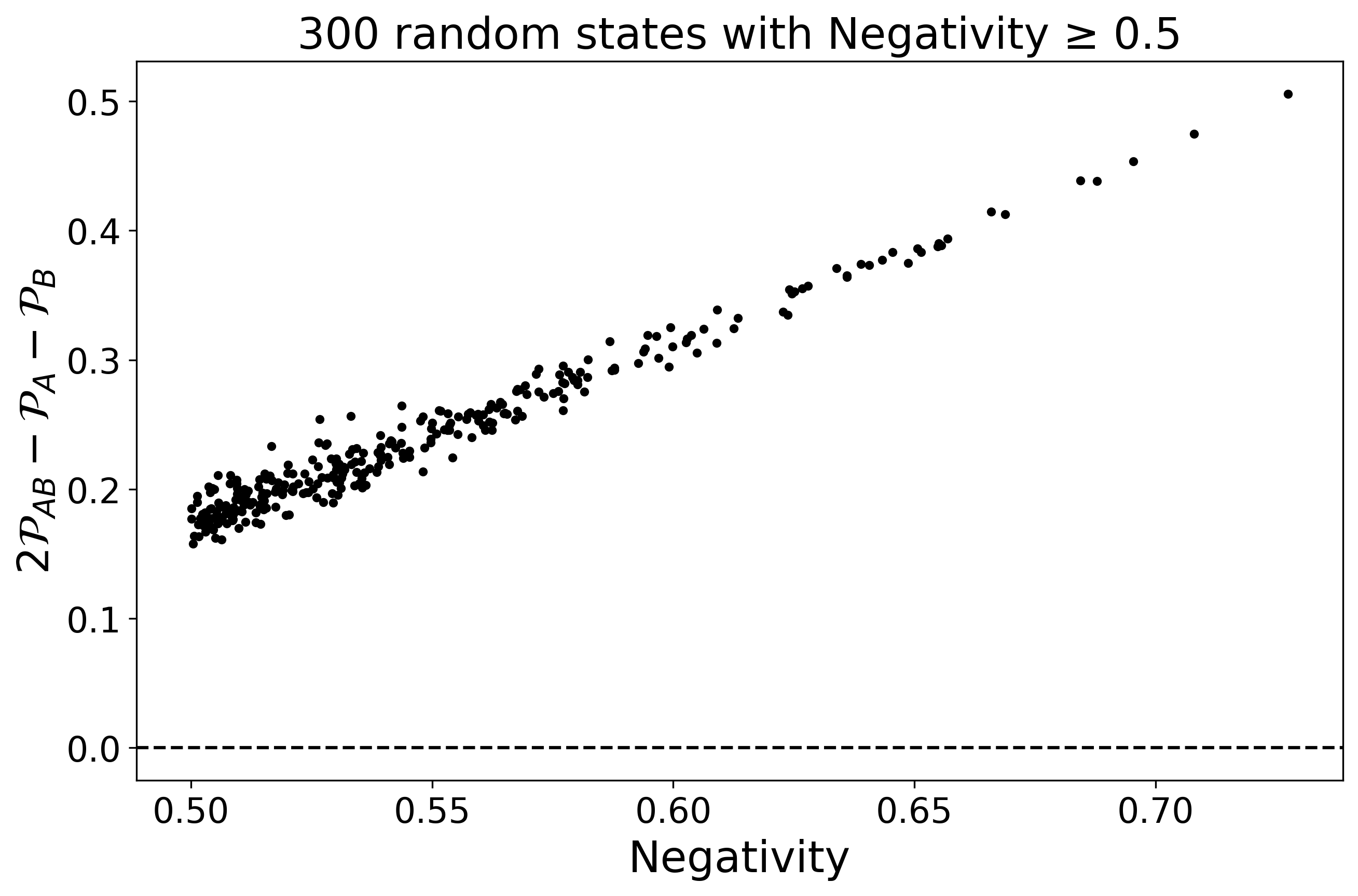}
        \caption{Generic random extracted two-qubit states with negativity $> 0.5$. The criterion never fails.}
        \label{Fig:pur_random_high_negativity_plot}
    \end{subfigure}
    \caption{$\|t\|^2 - 1$, expressed in terms of the global and reduced purities, as a function of the entanglement negativity for randomly extracted two-qubit states.}
    \label{Fig:combined_figure}
\end{figure}

For general two-qubit states, $\| t \|^{2}$ can no longer be interpreted as the distance from the MM state. For randomly generated states, one finds that the criterion detects entanglement reliably for values of the entanglement negativity larger than $1/2$, as illustrated in Fig.~\ref{Fig:combined_figure}.

\subsection{Higher dimensions}
\subsubsection{$4$-qubit generalized noisy GHZ state}
\begin{figure}[t!]
    \centering{\includegraphics[width=.45\textwidth]{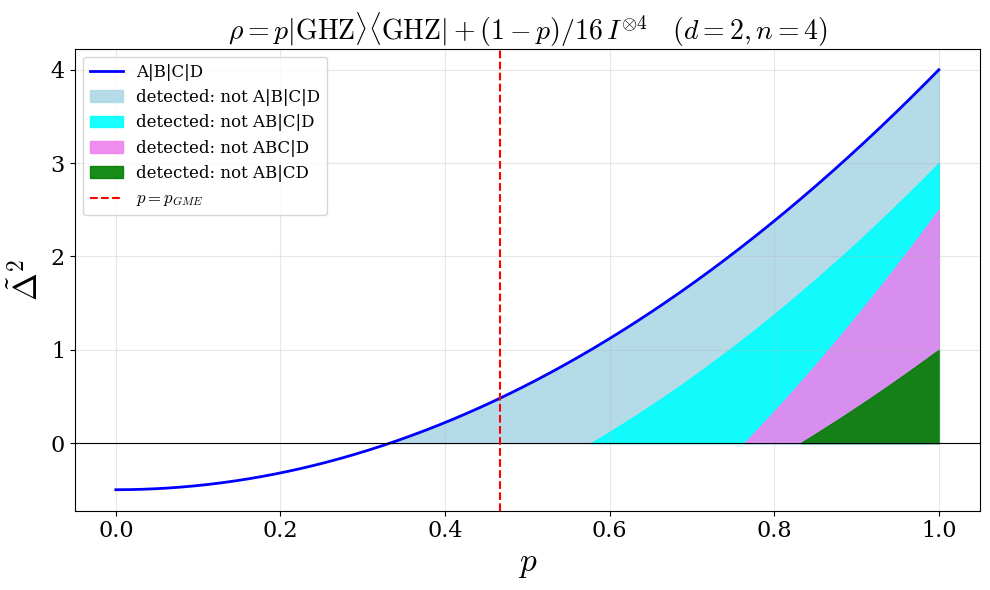}}
    \caption{$\widetilde{\Delta}^{2}$ in Eq.~\eqref{delta_tilde} as a function of $p$ for a noisy GHZ state of $n=4$ qubits, dimension $d=16$, and different $k$-partitions. The dotted vertical line identifies the region to the right of the threshold value $\bar{p}$, where the state features genuine multipartite entanglement.}
    \label{Fig:noisy_ghz}
\end{figure}

Noisy GHZ states of $n$ qubits are defined as the convex combination of GHZ states with the maximally mixed state in dimension $d = 2^{n}$:
\begin{equation}
    \rho^{(\mathrm{GHZ})} = p \ket{\mathrm{GHZ}}\bra{\mathrm{GHZ}} + (1-p)\frac{1}{d}\mathds{1}_{d} \, .
\end{equation}
Mixed GHZ states of $n$ qubits are particularly interesting because it is known that they are genuinely multipartite entangled (GME) if and only if \cite{Yamasaki2022}
\begin{equation}
        p > \frac{2^{n-1}-1}{2^{n }-1} \, .
\end{equation}
In Fig.~\ref{Fig:noisy_ghz} we report the performance of the criterion~\eqref{general_inequality_criterion} as a function of the classical probability $p$.

\subsubsection{n-qubit states with maximally disordered reductions}
For a system of $n$ qubits we wish to investigate the conditions for full separability. Putting $k=n$ at fixed $g$ in Eq.~\eqref{general_inequality_criterion}, we have
\begin{equation}
    \sum_{ \{A\}_{g} } d_{ \{A\}_{g} } \big( 1 - \mathcal{P}^{\{A\}_{g}} \big) = 2^{g} \sum_{i_{1} < \dots < i_{g}} \big( 1 - \mathcal{P}^{(A_{i_{1}}\dots A_{i_{g}})} \big)\, .
\end{equation}
Hence, the term
\begin{align}
   q &\equiv \sum_{g=1}^{n} (-1)^{n-g}2^{g} \sum_{i_{1}< \dots < i_{g}} 1 \notag \\
   &= \sum_{g=0}^{n} (-1)^{n-g} 2^{g}\binom{g}{n} - (-1)^{n}= 1 - (-1)^{n}
\end{align}
vanishes for odd $n$, while$q=2$ for even $n$. The criterion can then be recast as
\begin{equation}
    \sum_{g=1}^{n} (-1)^{n-g} \sum_{\{ A\}_{g} } 2^{g}\mathcal{P}^{ \{ A \}_{g}} - q > 0 \, .
\end{equation}
If all reduced states are maximally mixed (MM), one has $\mathcal{P}^{ \{ A\}_{g}} = 2^{-g}$ for any fixed $g < n$. Hence,
\begin{align}
    &2^{n} \mathcal{P}^{(A_{1}\dots A_{n})} + \sum_{g=1}^{n-1} \binom{i}{n}(-1)^{g-n} - q =  \notag \\
    & 2^{n} \mathcal{P}^{(A_{1}\dots A_{n})} - 2 > 0 \, .
\end{align}
Therefore, an $n$-qubit state is fully separable if its purity satisfies the following inequality:
\begin{equation}
    \mathcal{P}^{(A_{1}\dots A_{n})} > \frac{1}{2^{n-1}} \, .
\end{equation}
The above relation is a generalization of the one holding for two-qubit BD states. In general, a state with MM reductions is not fully separable if the global purity of the system is larger than twice the purity of the MM state.
Similar results apply to the $k$-separability of $n$-qubit states for $k < n$.

\subsubsection{Qudits}
Consider first a bipartite system of two qudits $A$, $B$ of dimension $d = d_{A}d_{B}$. In this case, the separability inequality~\eqref{general_inequality_criterion} reads
\begin{equation}
    d (1 - \mathcal{P}^{(AB)}) - d_{A} (1 - \mathcal{P}^{(A)}) - d_{B} (1 - \mathcal{P}^{(B)}) < 0 \, .
\end{equation}
If the two subsystems are of comparable dimensions $d_{A} \simeq d_{B}$ the efficacy of the criterion increases with decreasing $d$. Indeed, if the purities are of the same order of magnitude, the first term in the inequality is of order $\mathcal{O}(d_{A}^{2})$ and the inequality is never satisfied, except when $\mathcal{O}(1-\mathcal{P}^{(AB)}) =  \mathcal{O} \big( \frac{1}{d_{A}} \big)$, which implies $\mathcal{P}^{(AB)} \simeq 1$. For example, if we consider two qudits with $d_{A}=d_{B}=10$, the purity of the total state has to be at least $\mathcal{P^{(AB)}} \gtrsim 0.8$. The situation improves dramatically when $d_{A} \gg d_{B}$ (and vice-versa).

The range of validity of the criterion greatly increases when considering multipartite states of $n$ qudits, since there are binomial coefficients $\binom{l}{k}$ which can be of the same order of magnitude or larger than the system global dimension $d$ for a large enough number $k$ of parties, similarly to the situations already discussed for systems of $n$ qubits.

\section{Discussion and outlook}
\label{sec_conclusions}

In the present paper we have studied the inter-relations between quantum uncertainties, quantum state purities and quantum entropies, and the criteria for separability and entanglement in multipartite quantum systems. For any given $k$-partition of a multipartite system, we have derived an exact uncertainty relation valid for arbitrary quantum states. This relation expresses the total uncertainty of tensor products of $\mathrm{SU}(d_{A_{l}})$ observables, with $l=1,\dots,k$, solely in terms of the purities (or, equivalently, the associated two-Rényi entropies) of the global state and of its reduced states, revealing a fundamental conservation law relative to uncertainties and purities (or, equivalently, entropies).  

Exploiting the relation between the total uncertainty and the algebraic sum of the quantum state purities (global and reduced), we have obtained a general Robertson-Schr\"odinger uncertainty inequality for $\mathrm{SU}(N)$ observables, and we have computed explicitly the rest term that saturates the inequality for Pauli observables in single-qubit and T-diagonal two-qubit states, whenever the single or the two qubits are subsystems of any given arbitrary multipartite system..

Building on this framework, we have established a criterion for $k$-separability for a generic multipartite quantum system of arbitrary dimension. The criterion is entirely formulated in terms of purities (or, equivalently, in terms of two-Rényi entropies) and is expressed by an inequality whose violation provides a sufficient condition for entanglement. We have then showed how the criterion can be reformulated in terms of an inequality for the Hilbert-Schmidt (HS) squared norm $\|t\|^{2}$ of the correlation matrix $t$. In the particular case of $n$ qubits, we prove explicitly that our general criterion reduces to a previously known condition expressed in terms of the HS squared norm of $t$.
%The resulting formula allows one to express the Hilbert--Schmidt (HS) norm of the correlation matrix in terms of the %purity of the reduced state. 

Expressing the norm of the correlation matrix $t$ as the algebraic sum of quantum state purities (entropies) provides an experimentally accessible route to entanglement detection that applies not only to the $k$-separability criterion, but also to all correlation, entanglement, and nonlocality criteria based on the norm of $t$. Indeed, we have shown that the number of purity measurements needed to evaluate $\|t\|^{2}$ scales exponentially more efficiently with the size of the system compared to the number of direct measurements necessary to determine the individual elements of the correlation matrix. 

We have investigated the regimes in which the advantage of expressing $\|t\|^{2}$ in terms of the purities is most significant, and found that it is maximized as the mixedness of the global state increases, leading to a scaling $\|t\|^{2} \sim \mathcal{O}(1)$. Notably, the interval of high mixedness (low purity) is exactly the regime in which entanglement and nonlocality detections are most challenging, making the proposed approach especially relevant.

Moving to explicit examples, we have investigated the effectiveness of the criterion both analytically and numerically, identifying classes of states for which the criterion detects entanglement most efficiently. In the most elementary case of two-qubit states, we have established a simple relation with the violation of the CHSH inequality and we have provided a complete geometric characterization of Bell-diagonal states. 

For many-qubit systems and more in general for arbitrary systems of qudits, we have discussed other close connections with the identification of genuine multipartite entanglement as well as with the detection of Bell nonlocality in all those scenarios that involve two dichotomic measurements per party, showing that if a state is Bell-nonlocal in the given scenario with two measurements per party, then it is necessarily detected by our criterion. The converse does not hold: in general, the criterion can certify entanglement also for states that are local in the same fixed scenario.

Overall, our results shed some further light on the intricate interrelations between uncertainty, entanglement, nonlocality, and quantum state purities (entropies), providing a conceptual foundation and an experimentally friendly framework for the detection and characterization of quantum correlations.

Considering possible future research directions, one interesting goal would be the generalization of the total uncertainty and the corresponding uncertainty relations to classes of observables more general than $\mathrm{SU}(N)$, such as, e.g., orthogonal observables. 

Since our investigation has covered the case of observables forming a basis in the finite-dimensional Hilbert space of complex matrices, and thus also for finite-dimensional Hermitian operators, another very relevant extension of our work would be the generalization to continuous-variable systems, for which some foundational relations between entanglement and quantum state purities, global and reduced, have already been studied in the past~\cite{Adesso2003,Adesso2004}. 

From a broader perspective, we plan to investigate the possibility of combining our framework with other approaches to the characterization and detection of nonlocal quantum correlations in multipartite quantum systems, with the aim of obtaining more stringent criteria for $k$-separability and multipartite entanglement certification. Along these lines, another intriguing open question concerns the study of the possible relations connecting purity-based criteria for $k$-separability and multipartite entanglement certification with more general forms of quantum nonlocality and the violation of different classes of Bell inequalities~\cite{Brunner2014}. 

\begin{acknowledgments}  
The authors thank Lorenzo Leone for discussions. F.I. acknowledges funding from the Italian Ministry of University and Research, call PRIN PNRR 2022, project ``Harnessing topological phases for quantum technologies'', code P202253RLY, CUP D53D23016250001, and PNRR-NQSTI project ECoN: ``End-to-end long-distance entanglement in quantum networks", CUP J13C22000680006.
\end{acknowledgments}

\newpage

\onecolumngrid

\appendix

\section{Settings and notation}
\label{appendix_notation}
In order to familiarize the reader with the setting and the notation, worked out by examples, we consider the simplest possible case of two qubits. A two-qubit state can be expanded as
\begin{equation}
   \rho = \frac{1}{4}\left[ \sum_{\alpha_{1},\alpha_{2}=0}^{3} t_{\alpha_{1}\alpha_{2}} \sigma_{\alpha_{1}\alpha_{2}} \right]\, ,
\end{equation}
with $\sigma_{\alpha_{1}\alpha_{2}} = \sigma_{\alpha_{1}}\otimes \sigma_{\alpha_{2}}$, $\sigma_{0} = \mathds{1}_{2}$ and $\sigma_{i}$ for $i=1,2,3$ are the Pauli matrices and $t_{\alpha_{1} \alpha_{2}} = \text{Tr}[\rho \sigma_{\alpha_{1}\alpha_{2}}]$. The purity of $\rho$ is then
\begin{equation}
   \mathcal{P}^{(A_{1}A_{2})} = \frac{1}{4} \|T\|^{2}\, ,
\end{equation}
where $T$ is the ($4 \times 4$) matrix with elements $t_{\alpha_{1}\alpha_{2}}$ and $\|\cdot\|$ is the squared HS norm.
If we define the $3-$dimensional vectors $t^{(A_{1})}$ and $t^{(A_{2})}$ with elements $t_{i0}$ and  $t_{0i}$, respectively for $i=1,2,3$, and the $(3 \times 3)$ matrix $t^{(A_{1},A_{2})}$ with elements $t_{ij}$ for $i,j=1,2,3$, the purities of the subsystems are
\begin{equation}
    \mathcal{P}^{(A_{l})} = \frac{1}{2} \left[\| t^{(A_{l})} \|^{2}+1\right], \qquad l=1,2
\end{equation}
Since $\|T\|^{2} = 1 +  \| t^{(A_{1})} \|^{2} +  \| t^{(A_{2})} \|^{2} +  \|t^{(A_{1}A_{2})} \|^{2}$, we obtain
\begin{equation}
\label{bipartite_qubits_formula}
    \|t^{(A_{1}A_{2})} \|^{2} = 4\mathcal{P}^{(A_{1}A_{2})} - 2(\mathcal{P}^{(A_{1})} + \mathcal{P}^{(A_{2})}) + 1\, .
\end{equation}
The generalization of the last equation to $n-$qudit systems is given by equation~\eqref{general_formula}

\section{Proof of the equality Eq.~\eqref{general_formula}}
\label{appendix_proof}
We will prove Eq.~\eqref{general_formula} by induction. For a fixed $k$-partition, the induction basis for a single part $l$ is proven by tracing out state $\rho$ in Eq.~\eqref{general_do} with respect to the remaining the $k-1$ parts, yelding the reduced state
\begin{equation}
    \rho^{(A_{l})} = \frac{1}{d_{l}}\left( \mathds{1}_{d_{l}} + \sum_{i_{l}=1}^{D_{A_{l}}} t_{i_{l}} \sigma_{i_{l}}^{(l)} \right)\, .
\end{equation}
Hence, by using $\text{Tr}[\sigma_{i_{l}}^{(l)}\sigma_{j_{l}}^{(l)}] = d_{l} \delta_{ij}$ we obtain
\begin{equation}
   \| t^{(A_{l})} \|^{2} = d_{l} \mathcal{P}^{(A_{l})} - 1
\end{equation}
for any $l=1,\dots,k$.

Now, let us suppose that Eq.~\eqref{general_formula} is valid (for the fixed $k$-partition) for any integer $l<k$, that is
\begin{equation}
    \|t^{(A_{1}\dots A_{l})}\|^{2} =  \sum_{g=1}^{l} \sum_{ \{ A\}_{g} } (-1)^{l-g} d_{ \{A\}_{g} } \mathcal{P}^{\{ A\}_{g}} + (-1)^{l}
\end{equation}
for any combination $A_{1},\dots,A_{l} \in \{A \}$. Then, recalling the definition Eq.~\eqref{useful_eq} of the HS squared norm of the correlation tensor, we have:
\begin{equation}
        \|T\|^{2} = 1 + \sum_{i} \|t^{(A_{i})}\|^{2} + \sum_{i,j}\|t^{(A_{i},A_{j})}\|^{2}+ \dots + \| t^{(A_{1}\dots A_{k})}\|^{2} = 1 + \sum_{g=1}^{k} \sum_{\{A \}_{g}} \| t^{( \{ A \}_{g})} \|^{2} \, .
\end{equation}
We can use the above relation to extract $\| t^{(A_{1}\dots A_{k})}\|^{2}$, observing that since $\sum_{g=0}^{k} (-1)^{g} \binom{g}{k} = (1-1)^{k} = 0$, then the constant terms, independent of the purities, are of the form 
\begin{equation}
    \sum_{g=0}^{k-1} (-1)^{g} \binom{g}{k} = \sum_{g=0}^{k} (-1)^{g} \binom{g}{k} - (-1)^{k} = -(-1)^{k} \, .
\end{equation}

In the same way, considering a single part $l$ at fixed purity $\mathcal{P}^{(A_{l})}$, one has
\begin{equation}
    d_{l} \mathcal{P}^{(A_{l})} \sum_{g=0}^{k-2} (-1)^{g} \binom{g}{k-1} = -(-1)^{k-1} d_{l} \mathcal{P}^{A_{l}} \, ,
\end{equation}
and, in general, 
\begin{equation}
    d_{1}\dots d_{l}  \mathcal{P}^{(A_{1}\dots A_{l})} \sum_{g=0}^{k-l} (-1)^{g} \binom{g}{k-l+1} = -(-1)^{k-l}  d_{1}\dots d_{l} \mathcal{P}^{(A_{1}\dots A_{l})}\, .
\end{equation}
Finally, collecting all terms, we have
\begin{equation}
    \|t^{(A_{1}\dots A_{k})}\|^{2} =  \sum_{g=1}^{k} \sum_{ \{ A\}_{g} } (-1)^{k-g} d_{ \{A\}_{g} } \mathcal{P}^{\{ A\}_{g}} + (-1)^{k}\, ,
\end{equation}
which is our thesis $\Box$. 

\end{document}